\documentclass[12pt, english]{article}

\usepackage[bindingoffset=0.2in,%
            left=0.4in,right=0.6in,top=1in,bottom=1in]{geometry}
\usepackage{amsmath}
\usepackage{amssymb}
\usepackage{amsthm}
\usepackage{bbm}
\usepackage{bm}
\usepackage{enumitem}
\usepackage{graphicx}
\usepackage{natbib}
\usepackage{subcaption}
\usepackage{hyperref}
\usepackage{tabularx}
\usepackage{multirow}
\usepackage{tikz}
\usepackage{authblk}

\usetikzlibrary{arrows,fit,positioning,shapes}
\tikzstyle{na} = [baseline=-.5ex]
\tikzstyle{every picture}+=[remember picture]
\pgfdeclarelayer{bg}
\pgfsetlayers{bg,main}

\hypersetup{colorlinks=true,allcolors=blue}
\usepackage[labelformat=simple]{subcaption}

\setcitestyle{round}

\def\isarxiv{1}

\title{AdaptSPEC-X: Covariate Dependent Spectral Modeling of Multiple Nonstationary Time Series}
\ifdefined\isblinded
  \author{}
  \date{}
\else
  \author[1]{Michael Bertolacci}
  \author[2]{Ori Rosen}
  \author[3]{Edward Cripps}
  \author[4]{Sally Cripps}
  \affil[1]{University of Wollongong}
  \affil[2]{University of Texas at El Paso}
  \affil[3]{University of Western Australia}
  \affil[4]{University of Sydney}

  \date{}
\fi

\newcommand*{\bvec}{\bm{b}}
\newcommand*{\mvec}{\bm{m}}
\newcommand*{\qvec}{\bm{q}}
\newcommand*{\rvec}{\bm{r}}
\newcommand*{\uvec}{\bm{u}}
\newcommand*{\wvec}{\bm{w}}
\newcommand*{\xvec}{\bm{x}}
\newcommand*{\zvec}{\bm{z}}
\newcommand*{\betavec}{\bm{\beta}}
\newcommand*{\etavec}{\bm{\eta}}
\newcommand*{\lambdavec}{\bm{\lambda}}
\newcommand*{\tauvec}{\bm{\tau}}
\newcommand*{\muvec}{\bm{\mu}}
\newcommand*{\kappavec}{\bm{\kappa}}
\newcommand*{\xivec}{\bm{\xi}}

\newcommand*{\tmin}{t_\text{min}}
\newcommand*{\mis}{\text{mis}}
\newcommand*{\obs}{\text{obs}}
\newcommand*{\all}{\text{all}}
\newcommand*{\N}{\text{N}}

\newcommand*{\MSEmean}{\text{MSE}_\text{mean}}
\newcommand*{\MSEspec}{\text{MSE}_\text{spec}}

\newcommand*{\componentindex}{h}
\newcommand*{\siteindex}{j}
\newcommand*{\freqindex}{k}
\newcommand*{\timeindex}{t}
\newcommand*{\ncomponents}{H}
\newcommand*{\ntimes}{n}
\newcommand*{\nsites}{N}

\DeclareMathOperator{\logit}{logit}
\DeclareMathOperator{\diag}{diag}
\DeclareMathOperator{\ind}{\mathbbm{1}}

\makeatletter
\newcommand*\dashline{\rotatebox[origin=c]{90}{$\dabar@\dabar@\dabar@$}}
\makeatother

\ifdefined\isarxiv
  \author{}
  \date{}
\else
\fi

\begin{document}

\maketitle

\abstract{
  We present a method for the joint analysis of a panel of possibly nonstationary time series. The approach is Bayesian and uses a covariate-dependent infinite mixture model to incorporate multiple time series, with mixture components parameterized by a time varying mean and log spectrum. The mixture components are based on AdaptSPEC, a nonparametric model which adaptively divides the time series into an unknown number of segments and estimates the local log spectra by smoothing splines. We extend AdaptSPEC to handle missing values, a common feature of time series which can cause difficulties for nonparametric spectral methods. A second extension is to allow for a time varying mean. Covariates, assumed to be time-independent, are incorporated via the mixture weights using the logistic stick breaking process. The model can estimate time varying means and spectra at observed and unobserved covariate values, allowing for predictive inference. Estimation is performed by Markov chain Monte Carlo (MCMC) methods, combining data augmentation, reversible jump, and Riemann manifold Hamiltonian Monte Carlo techniques. We evaluate the methodology using simulated data, and describe applications to Australian rainfall data and measles incidence in the US. Software implementing the method proposed in this paper is available in the R package BayesSpec.
}


\section{Introduction}

When the available data are multiple time series thought to be realizations of nonstationary random processes, estimation of their time varying mean and spectrum offers insight into the behavior of the processes, including whether and how they have changed over time.
For example, an analysis of the spatial distribution of the frequency domain characteristics of
time series of rainfall for sites spanning a wide spatial field can quantify the cyclical variability of the underlying process, while allowing for nonstationarity can suggest ways in which the climate has changed over the observation period.
Extending the work of \citet{priestley1965}, \citet{dahlhaus1997} established an asymptotic framework for locally stationary processes in which the spectral density of the process is allowed to evolve over time. Key to these asymptotics is the notion that better estimates of the local spectrum for a single stochastic process come from observing the stochastic process at finer time intervals, rather than observing it over a longer time period. This is problematic for historic time series, e.g.,  rainfall, for which no further observations can be made. Joint modeling of multiple time series with similar or identical local spectra is one way to ameliorate this problem, as estimates for different time series can borrow strength from each other. If additional information is available, such as covariates, it should be incorporated into the model to improve estimation. This is the motivation behind the approach taken in this article.

In particular, this article presents methodology for analyzing a panel of possibly nonstationary time series using a covariate-dependent infinite mixture model, with mixture components parameterized by their time varying mean and spectrum. The mixture components are based on AdaptSPEC \citep{rosenetal2012}, which partitions a (centered) time series into an unknown but finite number of segments, estimating the spectral density within each segment by smoothing splines. As part of the proposed method, AdaptSPEC is extended to handle missing values, a common feature of time series which can cause difficulties for nonparametric spectral methods. A second extension is the incorporation of a time varying mean, which avoids having to de-mean (center) the time series as a preliminary step. The covariates, which are assumed to be time-independent, are incorporated via the mixture using the logistic stick breaking process (LSBP) of \citet{rigondurante2020}, where the log odds for each `stick break' are modeled using a thin plate spline Gaussian process over the covariates. The model is formulated in a Bayesian framework, where Markov chain Monte Carlo (MCMC) methods are used for parameter estimation and to deal with missing values. Specifically, as in AdaptSPEC, reversible jump MCMC (RJMCMC) is used to estimate the mixture component parameters, while the LSBP parameters are estimated via the P\'{o}lya-Gamma based latent variable expansion of \citet{rigondurante2020} \citep[see also][for the original latent variable expansion in the finite mixture case]{polsonetal2013}. The model and sampling scheme are capable of handling large panels, such as that of the measles application which has nearly 200,000 observations. In addition to estimating time varying spectra for each time series in the panel, the covariate-dependent mixture structure allows inference about the underlying process at unobserved covariate values, enabling predictive inference. For instance, in this work, we use longitude and latitude as covariates when modeling Australian rainfall data, and are able to infer the predictive time varying spectrum of the rainfall process at unobserved locations.

Many methods have been proposed for the spectral analysis of time series. As the focus of this paper is multiple time series, we provide background in the form of an overview of methods for nonstationary single time series. We then review methods for multiple time series, stationary or otherwise. This excludes methods for multivariate time series, and we refer readers to \citet{likrafty2018} for a review of past and recent work in this active research area.

Approaches to spectral estimation for a single nonstationary time series include fitting parametric time series models with time varying parameters \citep{kitagawagersch1996,dahlhaus1997,westetal1999,yangetal2016}, smoothing the log periodogram \citep{ombaoetal2001,guoetal2003}, dividing the time series into locally stationary segments \citep{adak1998,davisetal2006,rosenetal2009,rosenetal2012}, and using short time Fourier transforms \citep{yangzhou2020}. For a recent and extensive overview of methods for single nonstationary time series, see \citet{yangetal2016}. Most directly relevant to this paper, \citet{rosenetal2009} estimate the log of the spectral density using a Bayesian mixture of splines. The time series is partitioned into small sections, and it is assumed that the log spectral density within each partition is given by a mixture of smoothing splines. The mixture weights are assumed to be time varying. \citet{rosenetal2012} present the AdaptSPEC method, which avoids the fixed partitions of \citet{rosenetal2009}. AdaptSPEC partitions the time series into one or more variable length segments in an adaptive manner, modeling the log spectral density within each segment via a smoothing spline. This results in better estimates than those obtained from the method of \citet{rosenetal2009}. Furthermore, the method  can accommodate both slowly and abruptly varying processes, as well as identify stationary processes. AdaptSPEC forms the basis of our proposed model for multiple time series.

For multiple time series, \citet{kraftyetal2011} construct a covariate-dependent model for multiple stationary series in which the log spectrum has a mixed effects representation, where the effects are functions over the frequency domain. \citet{macaroprado2014} propose a Bayesian model for multiple stationary time series with a covariate-dependent spectral density composed as a sum of spectral densities corresponding to different levels of two or more factors. Following \citet{choudhurietal2004}, they model the spectral density associated with the factors via Bernstein-Dirichlet priors. \citet{kraftyetal2017} present a Bayesian model for stationary multivariate time series based on the work of \citet{rosenstoffer2007}, where multiple (multivariate) time series from different subjects are available, and subjects have an associated single covariate. \citet{bruceetal2018} present a method for multiple nonstationary time series with a single covariate that is referred to as conditional adaptive Bayesian spectrum analysis, which adaptively partitions both time- and covariate-space, modeling the spectrum within each partition by smoothing splines \citep[as in][]{rosenetal2012}. \citet{cadonnaetal2018} model multiple stationary time series using a Bayesian hierarchical model. The log-periodogram of a single stationary time series is modeled as a mixture of Gaussian distributions where the mixture weights and mean functions are frequency-dependent. The hierarchical model for multiple time series is constructed by setting the mean functions to be common to all time series while letting the weights vary between time series.

Adding to this literature, we present methodology, referred to as AdaptSPEC-X,  which combines four features: multiple time series, nonstationarity in both mean and spectrum, multiple covariates, and missing data. We demonstrate the method on simulated data, and show how it can be used to estimate the mean and spectra in two application areas: Australian rainfall data, and measles incidence in the United States. Software implementing AdaptSPEC-X is available in the R package BayesSpec\footnote{Available from the authors; latest CRAN version does not contain AdaptSPEC-X.}.

The paper proceeds as follows. Section~\ref{sec:model_single} describes AdaptSPEC, the model for single nonstationary time series forming the basis for the analysis of multiple nonstationary time series. AdaptSPEC-X, a covariate-dependent infinite mixture model, is presented in Section~\ref{sec:model_multiple}.  Section~\ref{sec:sampling_scheme} outlines the MCMC scheme used to estimate the model parameters. Section~\ref{sec:simulation_study} presents a study of the performance of AdaptSPEC-X on replicated simulated data from multiple time series. Section~\ref{sec:applications} describes the application areas. Australian rainfall is analyzed in Section~\ref{sec:applications_rainfall}, and measles incidence in the US is discussed in Section~\ref{sec:applications_measles}. Appendix~\ref{sec:conditional_distributions} provides details of the conditional distributions necessary for the sampling scheme, and Appendix~\ref{sec:appendix_lambda_derivation} expands on the covariance structure used to derive the conditional distribution of the missing values.

\section{Model for single time series}
\label{sec:model_single}

\citet{rosenetal2012} (henceforth RWS12) present the AdaptSPEC method for modeling single nonstationary time series which we summarize in this section. Let $\xvec = (x_1, \ldots, x_\ntimes)'$ be a time series of length $\ntimes$. Assume for ease of notation that $\ntimes$ is even, and suppose initially that $\xvec$ is a realization from a stationary process $\{ X_\timeindex \}$ with constant mean $\mu$ and a bounded positive spectral density $f(\omega)$ for $\omega\in (-\frac{1}{2},\frac{1}{2}]$. \citet{whittle1957} showed that, for large $\ntimes$, the likelihood of $\xvec$ can be approximated as
\begin{equation}
  p(\xvec \mid \mu, f)
    =
    \frac{1}{(2\pi)^{\ntimes / 2}}
    \frac{1}{\prod_{\freqindex = 1}^\ntimes f(\omega_\freqindex)^{1 / 2}}
    \exp\left\{
      -\frac{1}{2} \sum_{\freqindex = 1}^\ntimes \frac{I_\freqindex}{f(\omega_\freqindex)}
    \right\},
    \label{eqn:whittle_likelihood}
\end{equation}
where $\omega_\freqindex = \frac{\freqindex - 1}{\ntimes}$ for $\freqindex = 1, \ldots, \ntimes$ are the Fourier frequencies,
$I_\freqindex = |d_\freqindex|^2$ is the periodogram at $\omega_\freqindex$, and
\begin{equation}
  d_\freqindex = \frac{1}{\sqrt{\ntimes}}\sum_{\timeindex = 1}^\ntimes (x_\timeindex - \mu) e^{-2\pi i \omega_\freqindex (\timeindex - 1)}
  \label{eqn:dft}
\end{equation}
is the discrete Fourier transform (DFT) at $\omega_\freqindex$, with $i = \sqrt{-1}$. RWS12 follow \citet{wahba1990} by expressing $\log f$ as
\begin{equation}
  \log f(\omega) = \alpha_0 + h(\omega),
  \label{eqn:log_spectral_density_spline}
\end{equation}
and placing a smoothing spline prior on $h(\omega)$. Due to the evenness of $f(\omega)$ and the periodogram, $h(\omega)$ is modeled on the domain $\omega \in [0, 0.5]$, corresponding to the first $\frac{\ntimes}{2} + 1$ Fourier frequencies. This prior is expressed via a linear combination of $J$ basis functions, where $J < \frac{\ntimes}{2} + 1$ is chosen to balance prior flexibility and computational resources. See Appendix~\ref{sec:conditional_distribution_theta} and RWS12 for details.

Next we allow the underlying process $\{ X_\timeindex \}$ to be nonstationary. Let a time series consist of a number of segments, $m$, and let $\xi_{s, m}$ be the end of the $s$th segment, $s = 1, \ldots, m$, where $\xi_{0, m} = 0$ and $\xi_{m, m} = \ntimes$. Then we assume that $\{ X_\timeindex \}$ is piecewise stationary, with
\begin{equation}
  X_\timeindex = \sum_{s = 1}^m X_\timeindex^s \delta_{s, m}(\timeindex),
  \label{eqn:piecewise_process}
\end{equation}
where the processes $\{ X_\timeindex^s \}$ are independent and stationary with means $\mu_{s, m}$, spectral densities $f_{s,m}(\omega)$, and $\delta_{s, m}(\timeindex) = 1$ iff $\timeindex \in (\xi_{s - 1, m}, \xi_{s, m}]$. Consider a realization $\xvec$ from \eqref{eqn:piecewise_process}. RWS12 approximate the likelihood of $\xvec$ by
\begin{equation}
  g(\xvec \mid \Theta) = \prod_{s = 1}^m p(\xvec_{s, m} \mid \mu_{s, m}, f_{s, m}),
  \label{eqn:piecewise_whittle_likelihoods}
\end{equation}
where $\Theta = \{ m, \xivec_m, \muvec_m, f_{1, m}, \ldots, f_{m, m} \}$, $\xivec_m = (\xi_{1, m}, \ldots, \xi_{m, m})'$, $\muvec_m = (\mu_{1, m}, \ldots, \mu_{m, m})'$, $\xvec_{s, m} = \{ x_\timeindex : \delta_{s, m}(\timeindex) = 1 \}$ are the data for the $s$th segment, and $p(\xvec_{s, m} \mid \mu_{s, m}, f_{s, m})$ is the Whittle likelihood \eqref{eqn:whittle_likelihood}. The values of $m$, $\xivec_m$ and $f_{s, m}$ for $s = 1, \ldots, m$ are considered unknown and are assigned priors. For $f_{s, m}$, the prior in \eqref{eqn:log_spectral_density_spline} is used, while $m$ is given the discrete uniform prior between $1$ and $M$. (See RWS12 for the details of the prior on $\xivec_m$). RWS12 consider $\muvec_m$ to be known and equal to zero, but in this work we consider it unknown and assign to $\mu_{s, m}$ a uniform prior with support $\mu_- < \mu_{s, m} < \mu_+$. A minimum segment length $\tmin$ is set to ensure that there are sufficient time periods within each segment so that the Whittle likelihood approximation is appropriate.

RWS12 develop a reversible jump Markov chain Monte Carlo algorithm \citep{green1995} that samples from the posterior distribution of this model. They show that AdaptSPEC can handle both abruptly and slowly varying nonstationary time series, as well as identify whether a time series is stationary. AdaptSPEC forms the basis of our spectral estimation technique for multiple time series.

\section{Model for multiple time series}
\label{sec:model_multiple}

In this section we extend the model of Section~\ref{sec:model_single} to multiple time series. Suppose now that the stochastic process $\{ X_\timeindex \}$ has associated covariates $\uvec = (u_1, \ldots, u_P)'$. We model $\{ X_\timeindex \}$ with a covariate-dependent mixture structure
\begin{equation}
  \{ X_\timeindex \} \sim \sum_{\componentindex = 1}^\ncomponents \pi_\componentindex(\uvec) g_\componentindex(\{ X_\timeindex \} \mid \Theta_\componentindex),
  \label{eqn:single_mixture}
\end{equation}
where the mixture component distributions $g_\componentindex$ are instances of AdaptSPEC (Equation~\eqref{eqn:piecewise_whittle_likelihoods}) with parameters $\Theta_\componentindex = \{ m^\componentindex, \xivec^\componentindex_m, \muvec^\componentindex_m, f^\componentindex_{1, m}, \ldots, f^\componentindex_{m, m} \}$, $2 \leq \ncomponents \leq \infty$, and the mixture weights $\pi_\componentindex(\cdot)$ satisfy $0 \leq \pi_\componentindex(\cdot) \leq 1$ and $\sum_{\componentindex = 1}^\ncomponents \pi_\componentindex(\uvec) = 1$. Equation~\eqref{eqn:single_mixture} implies that $\{ X_\timeindex \}$'s distribution is determined by its covariates $\uvec$, which, importantly, do not vary with time. The purpose of the mixture structure in Equation~\eqref{eqn:single_mixture} is to induce covariate-dependence in a flexible, semi-parametric manner, and we do not use this structure to perform inference about clustering or the number of clusters among multiple time series.

Let $\{ \xvec_1, \ldots, \xvec_\nsites \}$ be a finite collection of $\nsites$ time series, of length $\ntimes$ each, where each time series $\xvec_\siteindex = (x_{1, \siteindex}, \ldots, x_{\ntimes, \siteindex})'$ has covariates $\uvec_\siteindex = (u_{1, \siteindex}, \ldots, u_{P, \siteindex})'$ for $\siteindex = 1, \ldots, \nsites$. Assuming independence conditional on $\pi_h(\cdot)$ and $\Theta_h$, it follows from Equation~\eqref{eqn:single_mixture} that the joint distribution of the collection is
\begin{equation}
  p(\xvec_1, \ldots, \xvec_\nsites) = \prod_{\siteindex = 1}^\nsites \sum_{\componentindex = 1}^\ncomponents \pi_\componentindex(\uvec_\siteindex) g_\componentindex(\xvec_\siteindex \mid \Theta_\componentindex).
  \label{eqn:multiple_mixture}
\end{equation}

\subsection{Model for mixture weights}

For the mixture weights $\pi_\componentindex(\uvec)$ in Equation~$\eqref{eqn:single_mixture}$, we use the logit stick-breaking prior (LSBP) developed by \citet{rigondurante2020}, according to which $\pi_\componentindex(\uvec)$ is given by
\begin{equation}
  \pi_\componentindex(\uvec) = v_\componentindex(\uvec) \prod_{\componentindex' = 1}^{\componentindex - 1} (1 - v_{\componentindex'}(\uvec)),
  \label{eqn:lsbp_weights}
\end{equation}
where $\logit v_\componentindex(\uvec) = w_\componentindex(\uvec)$, so that $w_\componentindex(\uvec)$ are the covariate-dependent log odds. This prior allows for $2 \leq \ncomponents \leq \infty$ mixture components, with $v_\ncomponents(\uvec) = 1$ when $\ncomponents < \infty$.

As described by \citet{rigondurante2020}, this construction can be interpreted via sequential (continuation-ratio) logits \citep{agresti2018}. Let $z_\siteindex \in \{ 1, 2, \ldots, \ncomponents \}$ be a latent indicator such that $(\xvec_\siteindex \mid z_\siteindex = \componentindex) \sim g_\componentindex(\xvec_\siteindex \mid \Theta_\componentindex)$. Then the LSBP can be represented in a generative manner as a sequence of decisions, where $p(z_\siteindex = 1) = v_1(\uvec_\siteindex) = (1 + \exp(-w_1(\uvec_\siteindex)))^{-1}$, $p(z_\siteindex = 2 \mid z_\siteindex > 1) = v_2(\uvec_\siteindex) = (1 + \exp(-w_2(\uvec_\siteindex)))^{-1}$, and so on, such that in general, $p(z_\siteindex = \componentindex \mid z_\siteindex > \componentindex - 1) = (1 + \exp(-w_\componentindex(\uvec_\siteindex)))^{-1}$.

The model in equations~\eqref{eqn:multiple_mixture} and \eqref{eqn:lsbp_weights} has a similar structure to the classic mixture of experts model \citep{jacobsetal1991}, in which the weights of a finite mixture depend on covariates through multinomial logits. Our motivation for choosing the LSBP (Equation~\eqref{eqn:lsbp_weights}) over multinomial logits is to obviate the choice of the number of mixture components. The model in equations~\eqref{eqn:multiple_mixture} and \eqref{eqn:lsbp_weights} is in principle an infinite mixture and so the question of the number of components becomes irrelevant. In practice, however, it is common to truncate the infinite representation at a suitably high but finite $K$ \citep{ishawaranjames2001}. The LSBP is analogous to the probit stick-breaking process \citep{chungdunson2009}, where a probit link function is used in place of the logit.

\subsection{Model for log odds}
\label{sec:model_log_odds}

We model the log odds by a Gaussian process (GP) prior $w_\componentindex(\uvec) \sim \text{GP}(\beta_{0, \componentindex} + \uvec' \betavec_\componentindex, \tau_\componentindex^2 \Omega(\cdot, \cdot))$, where $\beta_{0, \componentindex}$ is an intercept, $\betavec_\componentindex = (\beta_{1, \componentindex}, \ldots, \beta_{P, \componentindex})'$ is a vector of regression coefficients, $\tau_\componentindex ^ 2$ is a smoothing parameter, and $\Omega(\uvec, \uvec')$ is the covariance kernel constructed via the reproducing kernel Hilbert space defined by a $P$-dimensional thin-plate Gaussian process prior \citep[see][]{wood2013}.
For a finite collection of $\nsites$ time series $\{ \xvec_1, \ldots, \xvec_\nsites \}$, with associated covariates $\{ \uvec_1, \ldots, \uvec_\nsites\}$, the log odds vector, $\wvec_\componentindex = (w_\componentindex(\uvec_1), \ldots, w_\componentindex(\uvec_\nsites))'$, has a multivariate normal distribution
\begin{equation}
  \wvec_\componentindex \sim \N(\beta_{0, \componentindex} \bm{1}_\nsites + U \betavec_\componentindex, \tau_\componentindex^2 \Sigma_w),
  \label{eqn:gp_mvt_norm}
\end{equation}
where $\bm{1}_\nsites$ is an $\ntimes \times 1$ vectors of ones, $U = (\uvec_1, \ldots, \uvec_\nsites)'$ is an $\nsites \times P$ matrix, and $\Sigma_w$ is an $\nsites \times \nsites$ matrix whose $\siteindex_1 \siteindex_2$th entry is equal to $\Omega(\uvec_{\siteindex_1}, \uvec_{\siteindex_2})$. To facilitate the posterior sampling scheme in Section~\ref{sec:sampling_scheme}, we transform the problem via a basis expansion \citep{wood2013}.
Let $\Sigma_w = QDQ'$ be the eigenvalue decomposition of $\Sigma_w$, where $Q$ is an $\nsites \times \nsites$ orthogonal matrix whose columns are the eigenvectors of $\Sigma_w$, and $D$ is a diagonal matrix containing the eigenvalues of $\Sigma_w$. Define $U^\dagger = (\bm{1}_\nsites, U, QD^{1/2})$ by columnwise concatenation, let $\betavec^\mathrm{GP}_\componentindex$ be an $\nsites \times 1$ vector, and $\betavec^\dagger_\componentindex = (\beta_{0, \componentindex}, \betavec_\componentindex', \betavec^{\mathrm{GP}\prime}_\componentindex)'$. The first column of $U^\dagger$ is a vector of ones, the next $P$ columns of $U^\dagger$ (equal to $U$) are the original covariates, while the last $N$ columns (equal to $QD^{1/2}$) are basis functions, with associated coefficients $\betavec^\mathrm{GP}_\componentindex$. For computational convenience and parsimony, we truncate the basis expansion to the first $B < \nsites$ basis functions, so that $U^\dagger$ is $\nsites \times (P + B + 1)$ and $\betavec^\dagger_\componentindex$ is $(P + B + 1) \times 1$. Equation~\eqref{eqn:gp_mvt_norm} now takes the form
\begin{equation}
  \renewcommand{\arraystretch}{1.5}
  \begin{array}{lll}
    \wvec_\componentindex
    & = & U^\dagger \betavec^\dagger_\componentindex, \\
    \betavec^\mathrm{GP}_\componentindex
    & \sim & \N(\bm{0}_B, \tau^2_\componentindex I_B),
  \end{array}
  \renewcommand{\arraystretch}{1}
  \label{eqn:gp_mvt_norm_linear}
\end{equation}
where $\bm{0}_B$ is an $B \times 1$ vector of zeros, and $I_B$ is the $B \times B$ identity matrix. This basis expansion, combined with the interpretation via sequential logits given in the previous section, facilitate the development of the posterior sampling scheme presented in Section~\ref{sec:sampling_scheme}.

The expression for $v_\componentindex(\uvec_\siteindex)$ becomes $\logit v_\componentindex(\uvec_\siteindex) = \uvec^{\dagger\prime}_\siteindex \betavec^\dagger_\componentindex$, where $\uvec^{\dagger\prime}_\siteindex$ is the $\siteindex$th row of $U^\dagger$. The prior placed on $(\beta_{0, \componentindex}, \betavec_\componentindex')'$ is $\N(\muvec_\beta, \Sigma_\beta)$, where $\muvec_\beta$ is a $(P + 1) \times 1$ vector and $\Sigma_\beta$ is a $(P + 1) \times (P + 1)$ covariance matrix, so that
\[
  \betavec^\dagger_\componentindex \sim \N\left(
    \begin{pmatrix}
      \muvec_\beta \\
      \bm{0}_B
    \end{pmatrix},
    \begin{pmatrix} \Sigma_\beta & 0 \\ 0 & \tau^2_\componentindex I_B \end{pmatrix}
  \right).
\]
Finally, to complete the model specification, we assign $\tau_\componentindex$ a half-$t$ distribution \citep{gelman2006} with density
\[
  p(\tau_\componentindex) \propto \left( 1 + \frac{1}{\nu_\tau}\left(\frac{\tau_\componentindex}{A_\tau}\right)^2 \right)^{-(\nu_\tau + 1) / 2}\;\; ,\tau_\componentindex >0,
\]
where $A_\tau$ and $\nu_\tau$ are scale and degrees of freedom parameters, respectively. As described in Appendix~\ref{sec:conditional_distributions}, the half-$t$ distribution can be expressed as
a scale mixture of inverse Gamma distributions, which simplifies the sampling of $\tau_h$. For the results in this paper, we set $\muvec_\beta = \bm{0}$, $\Sigma_\beta = 100 I_{P + 1}$, $\nu_\tau = 3$ and $A_\tau = 10$.

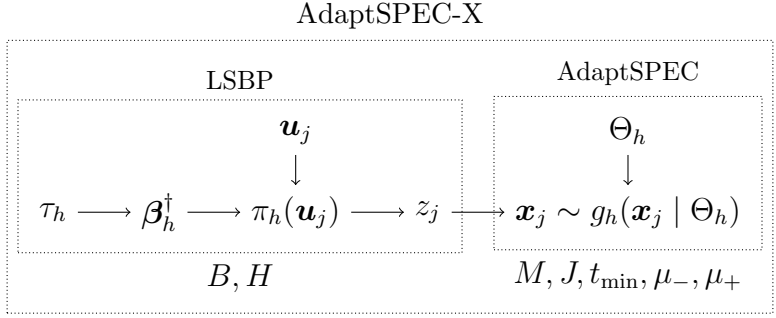
\begin{figure}
  \centering
  \begin{tikzpicture}[node distance=0.4cm and 0.7cm]
    \node (tau) {$\tau_\componentindex$};
    \node[right=of tau] (beta) {$\betavec^\dagger_\componentindex$};
    \node[right=of beta] (pi) {$\pi_\componentindex(\uvec_\siteindex)$};
    \node[above=of pi] (u) {$\uvec_\siteindex$};
    \node[right=of pi] (z) {$z_\siteindex$};
    \node[right=of z] (x) {$\xvec_\siteindex \sim g_\componentindex(\xvec_\siteindex \mid \Theta_\componentindex)$};
    \node[above=of x] (Theta) {$\Theta_\componentindex$};

    \node[draw,densely dotted,fit=(tau) (beta) (pi) (u) (z)] (lsbp) {};
    \node[draw,densely dotted,fit=(Theta) (x)] (adaptspec) {};

    \node[above] at (lsbp.north) (lsbp_title) {\footnotesize LSBP};
    \node[above] at (adaptspec.north) (adaptspec_title) {\footnotesize AdaptSPEC};

    \node[below] at (lsbp.south) (lsbp_hyper) {$B, H$};
    \node[below] at (adaptspec.south) (adaptspec_hyper) {$M, J, \tmin, \mu_-, \mu_+$};

    \node[draw,densely dotted,fit=(lsbp) (adaptspec) (lsbp_title) (adaptspec_title) (lsbp_hyper) (adaptspec_hyper)] (adaptspec_x) {};
    \node[above] at (adaptspec_x.north) (adaptspec_x_title) {\small AdaptSPEC-X};

    \path
      (tau)
        edge [->] (beta)
      (beta)
        edge [->] (pi)
      (u)
        edge [->] (pi)
      (pi)
        edge [->] (z)
      (z)
        edge [->] (x)
      (Theta)
        edge [->] (x);
  \end{tikzpicture}
  \caption{
    A graphical representation of AdaptSPEC-X. The bottom row lists the main hyperparameters.
  }
  \label{fig:graphical_model}
\end{figure}

Figure~\ref{fig:graphical_model} displays a graphical summary of the model in Equations~\eqref{eqn:piecewise_whittle_likelihoods} to \eqref{eqn:gp_mvt_norm_linear}, showing the dependence between the data, covariates, parameters, and hyperparameters.

\subsection{Missing values}
\label{sec:model_missing_values}

The model can accommodate missing values by exploiting the fact that the Whittle likelihood describes a multivariate normal distribution \citep{whittle1953}. As is shown below, this implies that the conditional distribution of the missing values is also multivariate normal, and we use this fact to accommodate missing values by integrating them out as part of the MCMC scheme in Section~\ref{sec:sampling_scheme}. For ease of exposition, in this section we first return to the case where $\xvec$ is a single stationary time series with spectral density $f$, then later describe how this is extended to multiple nonstationary time series. Define the $\ntimes \times \ntimes$ matrix $V$ with entries $V_{\timeindex\freqindex} = \frac{1}{\sqrt{\ntimes}} \exp\left( -2\pi i (\timeindex - 1)\omega_\freqindex \right)$ for $\timeindex = 1, \ldots, \ntimes$ and $\freqindex = 0, \ldots, \ntimes - 1$. It then follows that $V'(\xvec - \mu) = (d_1, \ldots, d_\ntimes)'$, where $d_\freqindex$ is given in \eqref{eqn:dft}. Noting that $V$ is a unitary matrix, \eqref{eqn:whittle_likelihood} may be rewritten as
\[
  p(\xvec \mid \mu, f)
    =
    \frac{1}{(2\pi)^{\ntimes / 2}}
    |R|^{1 / 2}
    \exp\left\{
      -\frac{1}{2} (\xvec - \mu)' V R V^* (\xvec - \mu)
    \right\},
    \label{eqn:whittle_likelihood_mvtnorm}
\]
where $R = \diag(\rvec)$, $\rvec = (1 / f(\omega_1), \ldots, 1 / f(\omega_\ntimes))'$, and $V^*$ is the conjugate transpose of $V$. The precision matrix $\Lambda = V R V^*$ is symmetric and circulant, and is thus defined by its first column, with entries $\Lambda_{\timeindex, 1} = \frac{1}{\ntimes} \sum_{\freqindex = 1}^\ntimes \frac{1}{f(\omega_\freqindex)} e^{-2\pi i (\timeindex - 1)\omega_\freqindex}$ (see Appendix~\ref{sec:appendix_lambda_derivation} for derivation). Suppose some values of $\xvec$ are missing, and write $\xvec = (\xvec_\mis', \xvec_\obs')'$, where $\xvec_\mis$ and $\xvec_\obs$ are the missing and observed values, respectively. From standard multivariate normal conditioning results,
\begin{equation}
  (\xvec_\mis \mid \xvec_\obs, \mu, f) \sim \N(\mu_{\mis \mid \obs}, \Lambda_{\mis \mid \obs}^{-1}),
  \label{eqn:missing_values_conditional}
\end{equation}
where
\begin{equation}
  \begin{split}
    \mu_{\mis \mid \obs}
    & = \mu - \Lambda_{\mis,\mis}^{-1} \Lambda_{\mis,\obs} (\xvec_\obs - \mu), \\
    \Lambda_{\mis \mid \obs}
    & = \Lambda_{\mis,\mis},
  \end{split}
  \label{eqn:missing_values_parameters}
\end{equation}
and the quantities in Equation~\eqref{eqn:missing_values_parameters} can be obtained from expressing $\Lambda$ as
\begin{equation}
  \Lambda = \begin{bmatrix}
    \Lambda_{\mis,\mis}  & \Lambda_{\mis,\obs} \\
    \Lambda_{\mis,\obs}' & \Lambda_{\obs,\obs}
  \end{bmatrix} = \begin{bmatrix}
    V_\mis R V_\mis^* & V_\mis R V_\obs^* \\
    V_\obs R V_\mis^* & V_\obs R V_\obs^* \\
  \end{bmatrix},
  \label{eqn:missing_values_lambda_decomposition}
\end{equation}
in which $V_\mis$ and $V_\obs$ are matrices made up of the rows of $V$ corresponding to the missing and observed times, respectively. The quantities in Equation~\eqref{eqn:missing_values_lambda_decomposition} can be computed efficiently by the fast Fourier transform. When simulating from Equation~\eqref{eqn:missing_values_conditional}, the most computationally intensive step is the inversion, $\Lambda_{\mis,\mis}^{-1}$, in Equation~\eqref{eqn:missing_values_parameters}. In a general framework for spectral estimation with stationary time series, \citet{guiness2019} describes computationally efficient methods for missing data imputation that could be used to simulate from Equation~\eqref{eqn:missing_values_conditional}. For this article we compute $\Lambda_{\mis,\mis}^{-1}$ the usual way, i.e., via its Cholesky decomposition.

Missing values may be accommodated in AdaptSPEC by partitioning the data in each segment, $\xvec_{s, m}$, into missing and observed times as above, and sampling from \eqref{eqn:missing_values_conditional} for each segment as part of the MCMC scheme. As shown in the next section, this is extended to the full AdaptSPEC-X model by conditioning on the latent indicators $\zvec$.

The Whittle likelihood is used in AdaptSPEC-X in two ways: first, as a nonparametric technique based on its asymptotic properties \citep[see Section~\ref{sec:model_single} and][]{rosenetal2012}, and second, in the assumption that the missing values follow a multivariate normal distribution. We justify the latter assumption in several ways. First, \citet{guiness2019}, who makes a similar assumption regarding missing values, found through both theory and numerical experiments that the assumption did not have a deleterious effect on spectral estimation. Second, it is hard to avoid making some assumption about the distribution of the missing values, and multivariate normality is arguably the most parsimonious choice, as it requires only the first two central moments to be specified (the minimum necessary to have mean and spectrum consistent with the observed data). Finally, as shown above and by \citet{guiness2019}, the assumption is computationally convenient.

\section{Sampling scheme}
\label{sec:sampling_scheme}

Define $\zvec = (z_1, \ldots, z_\nsites)'$, $\xvec^\all = (\xvec_1', \ldots, \xvec_\nsites')'$, $\betavec^\dagger = \{ \betavec^\dagger_1, \ldots, \betavec^\dagger_{\ncomponents - 1} \}$, $\Theta = \{ \Theta_1, \ldots, \Theta_\ncomponents \}$, and $\tauvec = (\tau_1, \ldots, \tau_{\ncomponents - 1})'$. Let $\xvec^\all = (\xvec^{\all\prime}_\mis \xvec^{\all\prime}_\obs)'$ be the decomposition of $\xvec^\all$ into missing and observed times, respectively. Values produced in steps 2, 3, 4, and 5 of the MCMC sampling scheme below are indicated by the superscript `$[l + 0.5]$', and are then used in a label swapping move in Step~\ref{enum:sampling_label_swap} to produce the $(l + 1)$th iteration. As described below, this move improves convergence of the following sampling scheme.
\begin{enumerate}[label=Step \arabic*.,ref=\arabic*,leftmargin=*]
  \item
    $(\xvec_\mis^{\all [l + 1]} \mid \betavec^{\dagger[l]}, \Theta^{[l]}, \zvec^{[l]}, \xvec^\all_\obs)$ as per Section~\ref{sec:model_missing_values}.
    \label{enum:sampling_missing}

  \item
    $(\Theta^{[l + 0.5]}_\componentindex \mid \xvec_\mis^{\all [l + 1]}, \zvec^{[l]}, \xvec^\all_\obs)$ for each $\componentindex = 1, \ldots, \ncomponents$. This step is potentially transdimensional, and uses the reversible-jump MCMC scheme of RWS12, with two modifications. The first one samples the segment means, $\muvec^\componentindex_m$, as described in Appendix~\ref{sec:sampling_scheme_means}. The second modification incorporates a Riemann manifold Hamiltonian Monte Carlo \citep[RMHMC for short, see][]{girolami2011} step to accelerate convergence, as described in Appendix~\ref{sec:sampling_scheme_rmhmc}.
    \label{enum:sampling_theta}

  \item
    $(\zvec^{[l + 0.5]} \mid \betavec^{\dagger[l]}, \Theta^{[l + 0.5]}, \xvec_\mis^{\all [l + 1]}, \xvec^\all_\obs)$.
    \label{enum:sampling_zvec}

  \item
    $(\betavec^{\dagger[l + 0.5]}_\componentindex \mid \tauvec^{[l]}, \zvec^{[l + 0.5]})$ for each $\componentindex = 1, \ldots, \ncomponents$. This uses the Polya-Gamma data augmentation scheme developed by \citet{polsonetal2013}, as applied to the LSBP by \citet{rigondurante2020}.
    \label{enum:sampling_beta}

  \item
    $(\tau^{[l + 0.5]}_\componentindex \mid \betavec^{\dagger[l + 0.5]})$ for each $\componentindex = 1, \ldots, \ncomponents$.
    \label{enum:sampling_tau}

  \item
    $(\Theta^{[l + 1]}, \zvec^{[l + 1]}, \betavec^{\dagger[l + 1]}, \tauvec^{[l + 1]} \mid \xvec_\mis^{\all [l + 1]}, \xvec^\all_\obs)$ using a label swapping step, described below.
    \label{enum:sampling_label_swap}
\end{enumerate}

The details of steps~\ref{enum:sampling_theta}, \ref{enum:sampling_zvec}, \ref{enum:sampling_beta}, and \ref{enum:sampling_tau} are presented in Appendix~\ref{sec:conditional_distributions}, while Step~\ref{enum:sampling_missing}, for $\xvec_\mis$, is described in Section~\ref{sec:model_missing_values}. For Step~\ref{enum:sampling_label_swap}, we adapt a label swapping move from \citet{hastieetal2015}, who find that it improves convergence in the context of MCMC samplers for Dirichlet process mixture models. The label swapping step is composed of the following substeps:

\begin{enumerate}[label=Step 6\alph*.,ref=6\alph*,leftmargin=*]
  \item
    Pick uniformly at random components $\componentindex_1, \componentindex_2 \in \{ 1, \ldots, \ncomponents \}$, $\componentindex_1 < \componentindex_2$, to swap.

  \item
    Construct proposal component indicators $\zvec^\text{swap}$ and $\Theta^\text{swap}$ such that
    \begin{align*}
      z^\text{swap}_\siteindex = \begin{cases}
        \componentindex_2 & \text{if } z^{[l + 0.5]}_\siteindex = \componentindex_1, \\[0.1cm]
        \componentindex_1 & \text{if } z^{[l + 0.5]}_\siteindex = \componentindex_2, \\[0.1cm]
        z^{[l + 0.5]}_\siteindex & \text{otherwise,}
      \end{cases}
      &
      \qquad \Theta^\text{swap}_\componentindex = \begin{cases}
        \Theta^{[l + 0.5]}_{\componentindex_2} & \text{if } \componentindex = \componentindex_1, \\[0.1cm]
        \Theta^{[l + 0.5]}_{\componentindex_1} & \text{if } \componentindex = \componentindex_2, \\[0.1cm]
        \Theta^{[l + 0.5]}_{\componentindex} & \text{otherwise.}
      \end{cases}
    \end{align*}
    \label{enum:label_swap_swap}

  \item
    Construct proposal $\tauvec^\text{swap}$ by setting
    \[
      \tau^\text{swap}_\componentindex = \begin{cases}
        \tau^{[l + 0.5]}_{\componentindex_2} & \text{if } \componentindex = \componentindex_1 \\[0.1cm]
        \tau^{[l + 0.5]}_{\componentindex_1} & \text{if } \componentindex = \componentindex_2 \\[0.1cm]
        \tau^{[l + 0.5]}_{\componentindex} & \text{otherwise,}
      \end{cases}
    \]
    and sample proposal $\betavec^{\dagger\text{swap}}_{\componentindex_1}, \betavec^{\dagger\text{swap}}_{\componentindex_2}$ from
    \[
      q(\betavec^{\dagger\text{swap}}_\componentindex \mid \zvec^{[l + 0.5]}, \tauvec^{[l + 0.5]}) \sim \N(
        \mu^\text{mode}_\componentindex,
        \Sigma^\text{mode}_\componentindex
      ),
    \]
    where $\mu^\text{mode}_\componentindex$ and $\Sigma^\text{mode}_\componentindex$ are the mode and the negative inverse of the Hessian, respectively, of $\log p(\betavec^{\dagger\text{swap}}_\componentindex \mid \zvec^\text{swap}, \tauvec^\text{swap})$.
    \label{enum:label_swap_tau_beta}

  \item
    Accept the swap with probability equal to the Metropolis-Hastings ratio
    \begin{align*}
      & \min\Bigg\{
        1,
        \frac{
          p(\betavec^{\dagger\text{swap}}, \zvec^\text{swap}, \Theta^\text{swap}, \tauvec^\text{swap} \mid \xvec_\mis^{\all [l + 1]}, \xvec^\all_\obs)
        }{
          p(\betavec^{\dagger[l + 0.5]}, \zvec^{[l + 0.5]}, \Theta^{[l + 0.5]}, \tauvec^{[l + 0.5]} \mid \xvec_\mis^{\all [l + 1]}, \xvec^\all_\obs)
        } \\
        & \qquad \qquad
        \frac{
          q(\betavec^{\dagger[l + 0.5]}_{\componentindex_1} \mid \zvec^\text{swap}, \tauvec^\text{swap})
        }{
          q(\betavec^{\dagger\text{swap}}_{\componentindex_1} \mid \zvec^{[l + 0.5]}, \tauvec^{[l + 0.5]})
        }
        \frac{
          q(\betavec^{\dagger[l + 0.5]}_{\componentindex_2} \mid \zvec^\text{swap}, \tauvec^\text{swap})
        }{
          q(\betavec^{\dagger\text{swap}}_{\componentindex_2} \mid \zvec^{[l + 0.5]}, \tauvec^{[l + 0.5]})
        }
      \Bigg\}.
    \end{align*}
    If accepted, set $\betavec^{\dagger[l + 1]}, \zvec^{[l + 1]}, \Theta^{[l + 1]}$ and $\tauvec^{[l + 1]}$ equal to $\betavec^{\dagger\text{swap}}, \zvec^\text{swap}, \Theta^\text{swap}$ and $\tauvec^\text{swap}$, respectively. Otherwise, $\betavec^{\dagger[l + 1]}, \zvec^{[l + 1]}, \Theta^{[l + 1]}$ and $\tauvec^{[l + 1]}$ are set to $\betavec^{\dagger[l + 0.5]}, \zvec^{[l + 0.5]}, \Theta^{[l + 0.5]}$ and $\tauvec^{[l + 0.5]}$, respectively.
\end{enumerate}

In Step~\ref{enum:label_swap_swap}, the labels of the component indicators for the chosen pair $\componentindex_1, \componentindex_2$ are swapped, as are the corresponding mixture component parameters $\Theta_\componentindex$, leaving the likelihood unchanged. Step~\ref{enum:label_swap_tau_beta} swaps the smoothing spline parameters $\tau_{\componentindex_1}, \tau_{\componentindex_2}$ and samples new values of $\betavec^\dagger_\componentindex$ from a normal approximation centered on its conditional mode. The latter new values are necessary because due to the sequential nature of the LSBP, merely swapping the values of $\betavec^\dagger_{\componentindex_1}$ and $\betavec^\dagger_{\componentindex_2}$ is unlikely to result in an acceptable proposal.

AdaptSPEC-X is implemented in the latest version of the R package BayesSpec.\footnote{Available from the authors; latest CRAN version does not contain AdaptSPEC-X.} The implementation is in R and C++, and can take advantage of multiple processor cores to reduce the running time of the analysis.

\section{Simulation study}
\label{sec:simulation_study}

\begin{figure}
  \centering
  \begin{subfigure}[t]{0.49\textwidth}
    \centering
    \includegraphics{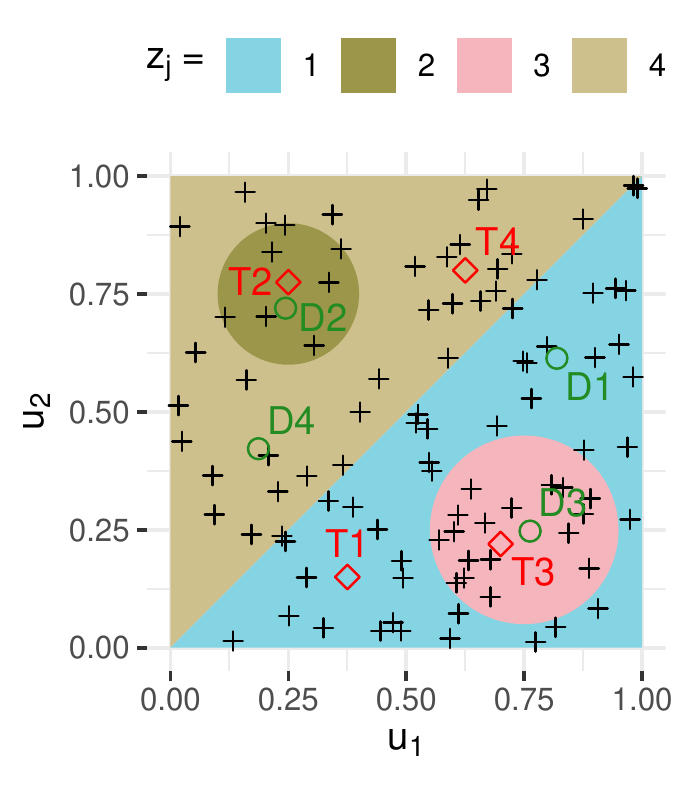}
    \caption{}
    \label{fig:multiple_simulation_study_true_categories}
  \end{subfigure}
  \begin{subfigure}[t]{0.49\textwidth}
    \centering
    \includegraphics{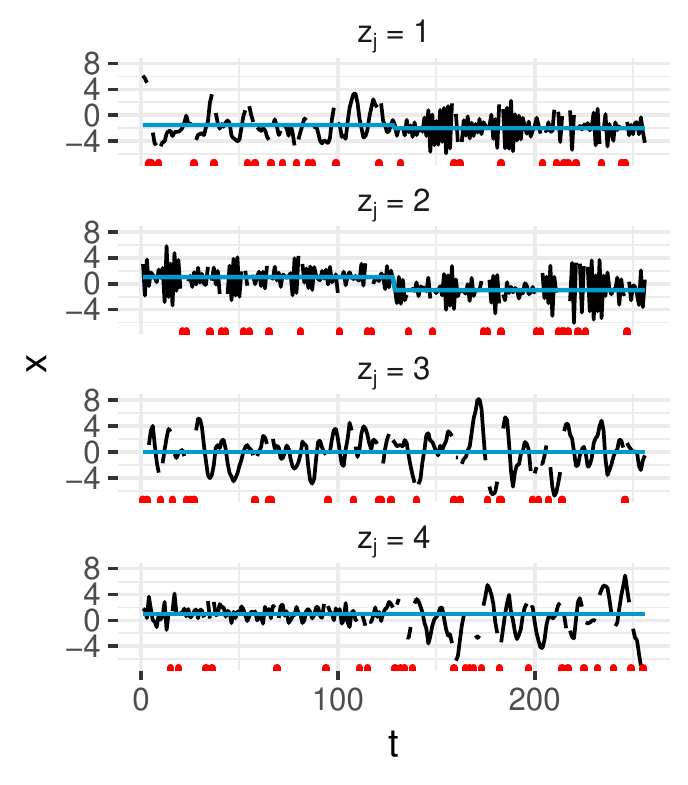}
    \caption{}
    \label{fig:multiple_simulation_study_ts}
  \end{subfigure}

  \caption{
    (a) Underlying surface mapping $\uvec_\siteindex$ to $z_\siteindex$ for Process~\eqref{eqn:multiple_simulation_study_model}, where the $100$ sampled locations are shown as crosses. The color of a region indicates the corresponding cluster. Four of the crosses are used as examples in the paper and are colored green and labeled D1 to D4. Four test points labeled T1 to T4 are shown with red diamonds.
    (b) Example realizations from Process~\eqref{eqn:multiple_simulation_study_model} corresponding to $z_\siteindex = 1$ at the top through $z_\siteindex = 4$ at the bottom. Black lines give the values of the time series, blue lines the underlying time varying mean $\mu_{z_\siteindex, \timeindex}$, and red ticks on the bottom axis mark missing values.
  }
\end{figure}

We now demonstrate AdaptSPEC-X using replicated simulated data including covariates and multiple time series with known time varying mean and spectrum. We are interested in the model's ability to recover the means and spectra at both observed and unobserved covariate values. Let $U = (\uvec_1, \ldots, \uvec_{100})'$ be a $100 \times 2$ design matrix corresponding to $N = 100$ subjects, each with two covariates, where the $\uvec_\siteindex$ are sampled uniformly from $[0, 1] \times [0, 1]$. Each $\uvec_\siteindex$ is mapped deterministically to $z_\siteindex \in \{ 1, 2, 3, 4 \}$, according to the plot shown in Figure~\ref{fig:multiple_simulation_study_true_categories}, which also includes the locations of the $100$ sampled points, denoted by crosses. Four locations are chosen as example time series, marked in green circles, and labeled D1 through to D4 (corresponding to $z_\siteindex = 1$ through $z_\siteindex = 4$, respectively). Four more locations are marked in red diamonds labeled T1 through to T4 (again for $z_\siteindex = 1$ to $4$). These values of $\uvec$ have no corresponding time series and are used as test points to evaluate the predictive inferences. The four different regions correspond to four different data generating processes. Each time series $\xvec_\siteindex$, within region $z_\siteindex$, $\siteindex = 1, \ldots, 100$, is a realization of length $n = 256$ from
\begin{equation}
  (x_{\siteindex, \timeindex} - \mu_{z_\siteindex, \timeindex})
  =
    \phi_{z_\siteindex, 1, \timeindex} (x_{\siteindex, \timeindex - 1} - \mu_{z_\siteindex, \timeindex - 1})
    + \phi_{z_\siteindex, 2, \timeindex} (x_{\siteindex, \timeindex - 2} - \mu_{z_\siteindex, \timeindex - 2})
    + \epsilon_{\siteindex, \timeindex},  \\
  \label{eqn:multiple_simulation_study_model}
\end{equation}
where $\epsilon_{\siteindex, \timeindex} \sim \N(0, 1)$, and the values of $\mu_{z_\siteindex, \timeindex}$ and $\phi_{z_\siteindex, p, \timeindex}$ are given in the following table:
\begin{center}
  \begin{tabular}{l|r|rr||r|rr}
                       & \multicolumn{3}{c||}{$t \leq 128$} & \multicolumn{3}{c}{$t > 128$} \\
                       & $\mu_{z_\siteindex, t}$ & $\phi_{z_\siteindex, 1, t}$ & $\phi_{z_\siteindex, 2, t}$
                       & $\mu_{z_\siteindex, t}$ & $\phi_{z_\siteindex, 1, t}$ & $\phi_{z_\siteindex, 2, t}$ \\ \hline
    $z_\siteindex = 1$ & -1.5 & 1.5  & -0.75
                       & -2   & -0.8 & 0 \\
    $z_\siteindex = 2$ & 1    & -0.8 & 0
                       & -1   & -0.8 & 0 \\
    $z_\siteindex = 3$ & 0    & 1.5  & -0.75
                       & 0    & 1.5  & -0.75 \\
    $z_\siteindex = 4$ & 1    & 0.2  & 0
                       & 1    & 1.5  & -0.75 \\
  \end{tabular}
\end{center}

Thus, time series with $z_\siteindex = 1$ have two segments with different means and different spectra, those with $z_\siteindex = 2$ have two segments with different means but with the same spectra, time series with $z_\siteindex = 3$ have only one stationary segment, and those with $z_\siteindex = 4$ have two segments with the same mean but with different spectra. In each time series, 10\% of the times are set as missing. Figure~\ref{fig:multiple_simulation_study_ts} displays example realizations from Process~\eqref{eqn:multiple_simulation_study_model} for $z_j = 1, 2, 3$ and $4$, showing the time series values, underlying time varying mean, and the times at which values are missing.

\begin{figure}
  \centering

  \includegraphics{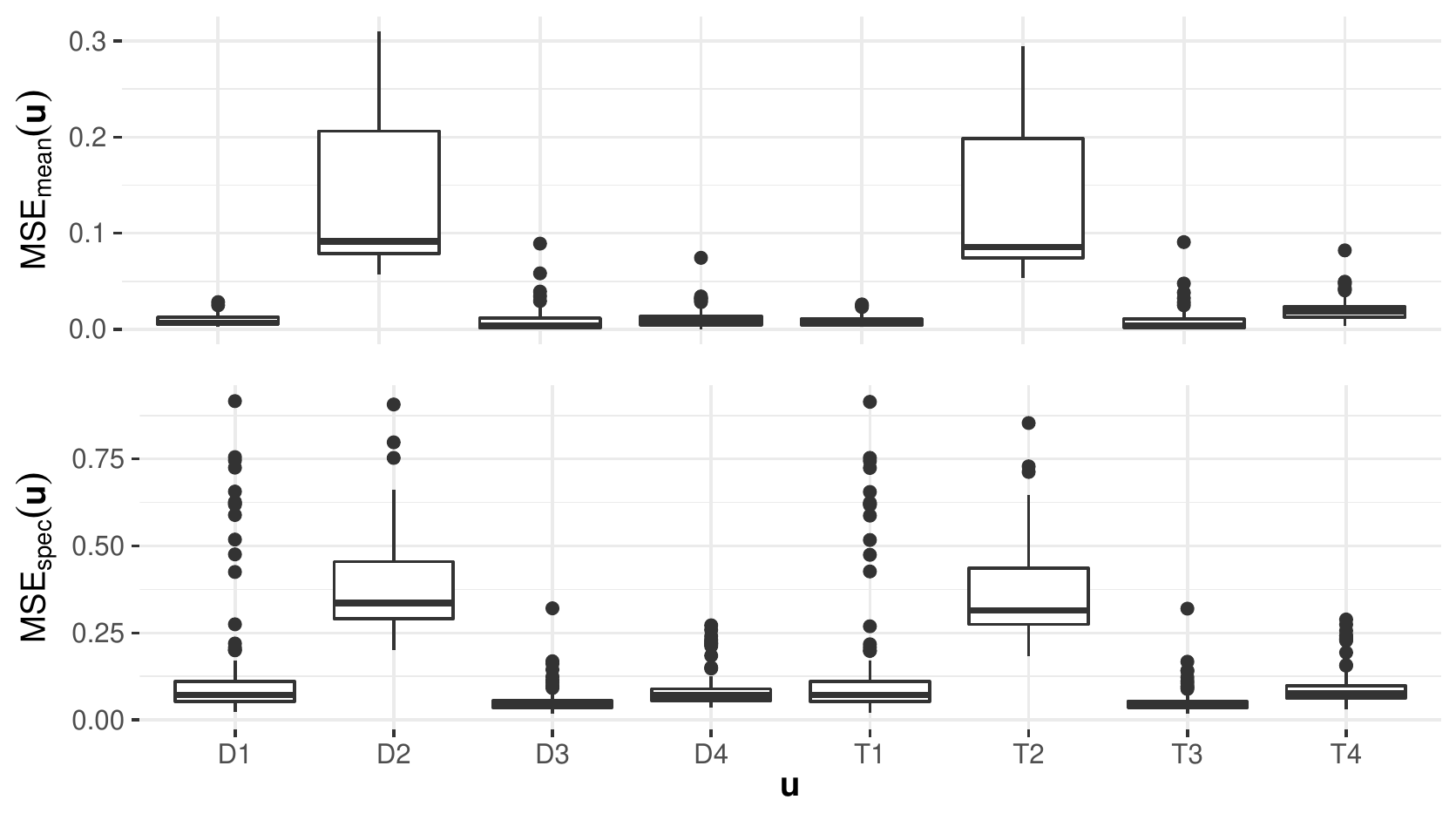}

  \caption{MSE across 100 replications for estimates of mean (top) and spectrum (bottom) for Process~\eqref{eqn:multiple_simulation_study_model} at the observed $\uvec =$ D1--D4 and the unobserved $\uvec =$ T1--T4 from left to right, respectively.}
  \label{fig:multiple_simulation_study_mse_boxplot}
\end{figure}

\begin{figure}
  \centering
  \includegraphics{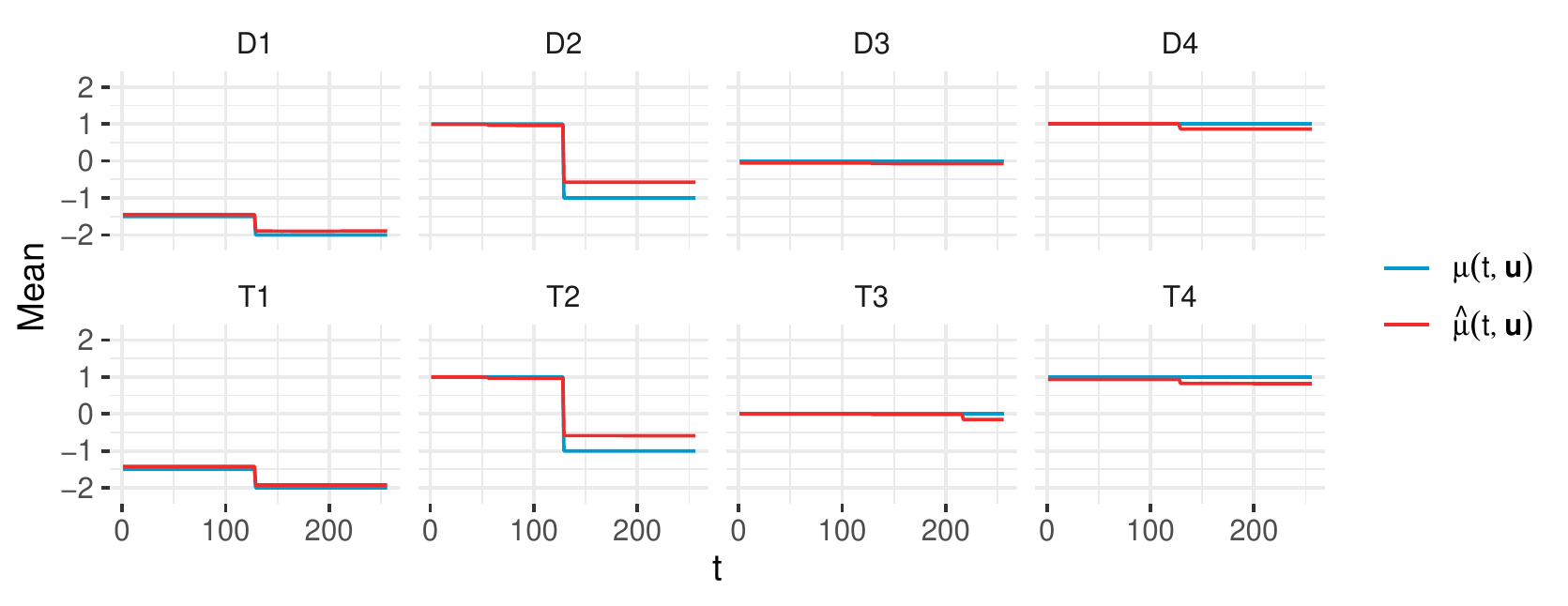}

  \caption{Estimated mean $\hat{\mu}(\timeindex, \uvec)$ corresponding to the median $\MSEmean(\uvec)$ (red) and true mean $\mu(\timeindex, \uvec)$ (blue) for Process~\eqref{eqn:multiple_simulation_study_model}. The first row shows the estimates for $\uvec =$ D1--D4 from left to right, respectively, while the second row shows the estimates for test points $\uvec =$ T1--T4.}
  \label{fig:multiple_simulation_study_tvm}
\end{figure}

\begin{figure}
  \centering
  \includegraphics{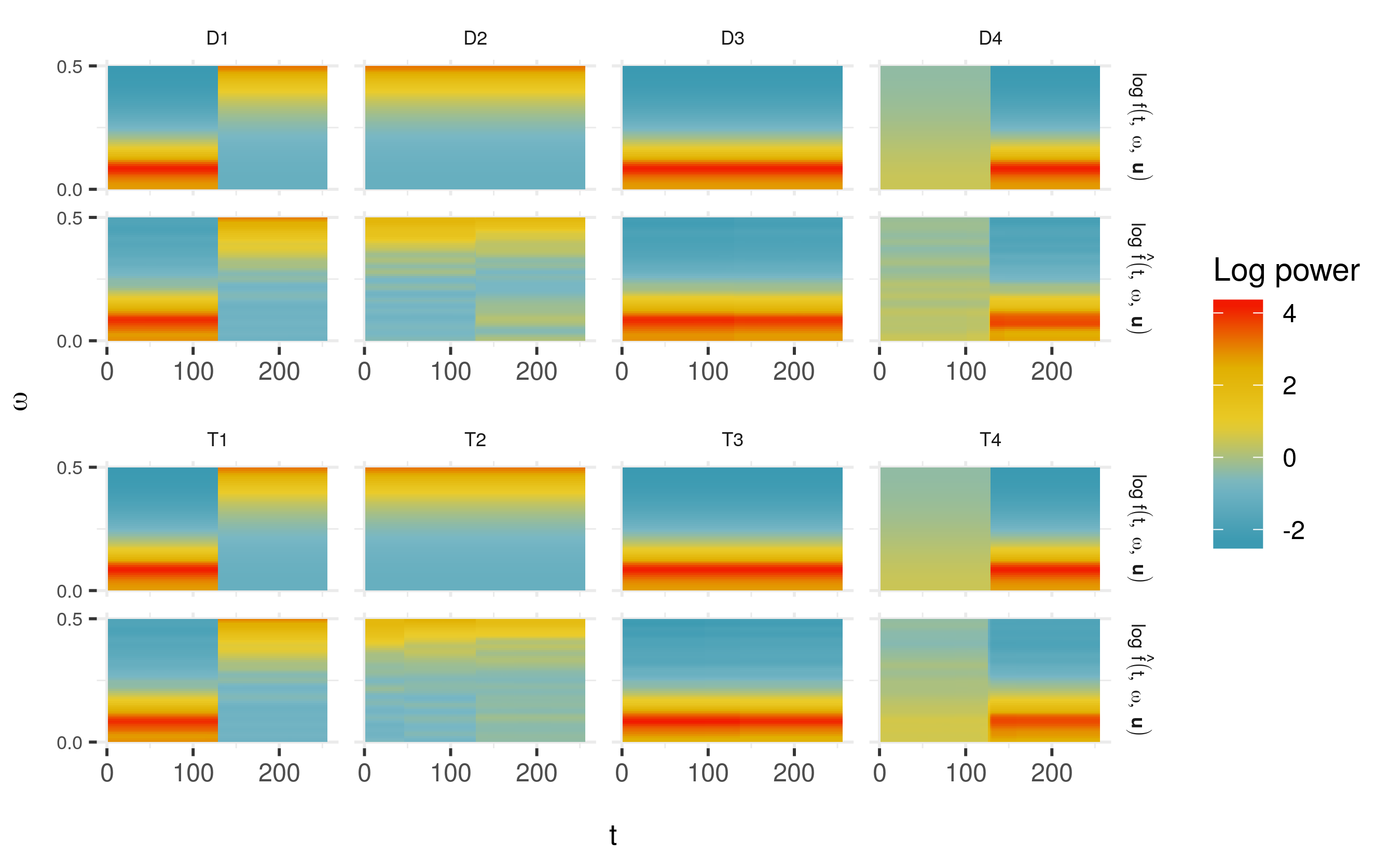}

  \caption{Estimated time varying log spectra $\log \hat{f}(\timeindex, \omega, \uvec)$ corresponding to the median $\MSEspec(\uvec)$ and true time varying log spectra $\log f(\timeindex, \omega, \uvec)$ for Process~\eqref{eqn:multiple_simulation_study_model}. The first row shows $\log f(\timeindex, \omega, \uvec)$ for $\uvec =$ D1--D4 from left to right, respectively, while the second row shows the estimates $\log \hat{f}(\timeindex, \omega, \uvec)$. The third and fourth rows display the analogous quantities for the test points $\uvec =$ T1--T4.}
  \label{fig:multiple_simulation_study_tvs}
\end{figure}

We sample 100 replicates from Process~$\eqref{eqn:multiple_simulation_study_model}$, and to each fit AdaptSPEC-X using the MCMC sampling scheme of Section~\ref{sec:sampling_scheme}. We run 50,000 iterations of the MCMC scheme, where the first 10,000 are discarded as burn-in. Each mixture component has $M = 4$, as the maximum number of segments, $\tmin = 40$, as the minimum segment length, $J = 25$, as the number of basis functions for the smoothing spline prior on the log spectra, and $(\mu_-, \mu_+) = (-10, 10)$, as the support of the prior on $\mu^\componentindex_{s, m}$. The LSBP is truncated at $\ncomponents = 25$ components, and has $B = 10$ basis functions. To assess the quality of the estimated time varying mean, we define the mean squared error (MSE) for the mean as
\begin{equation}
  \MSEmean(\uvec) = \frac{1}{\ntimes} \sum_{\timeindex = 1}^\ntimes \left[ \hat{\mu}(\timeindex, \uvec) - \mu(\timeindex, \uvec) \right]^2,
\end{equation}
where $\hat{\mu}(\timeindex, \uvec)$ is the estimate of $\mu(\timeindex, \uvec)$, the true time varying mean at covariates $\uvec$. Similarly, we define the MSE for the spectrum as
\begin{equation}
  \MSEspec(\uvec) =
  \frac{1}{\ntimes} \frac{1}{\freqindex_\text{max}}
  \sum_{\timeindex = 1}^\ntimes
  \sum_{\freqindex = 1}^{\freqindex_\text{max}} \left[
    \log \hat{f}\left( \timeindex, \frac{\freqindex - 1}{2\freqindex_\text{max} - 2}, \uvec \right)
    - \log f\left( \timeindex, \frac{\freqindex - 1}{2\freqindex_\text{max} - 2}, \uvec \right)
  \right] ^ 2,
\end{equation}

where $\freqindex_\text{max} = 128$, and $\log \hat{f}(\timeindex, \omega, \uvec)$ is the estimate of $\log f(\timeindex, \omega, \uvec)$, the true time varying log spectral density at location $\uvec$. Figure~\ref{fig:multiple_simulation_study_mse_boxplot} presents boxplots of $\MSEmean$ (top) and $\MSEspec$ (bottom), at each observed location, D1--D4, and unobserved test location, T1--T4, from left to right, respectively. The median $\MSEmean$ is less than 0.02 at all covariate values except for D2 and T2, for which it is 0.09. Similarly, the median $\MSEspec$ is less than 0.08 at all locations except for D2 and T2, for which it is 0.34 and 0.32, respectively. Estimates of the time varying mean and spectrum corresponding to the median MSE values are shown in figures~\ref{fig:multiple_simulation_study_tvm} and \ref{fig:multiple_simulation_study_tvs}, respectively. These qualitatively match the $\MSEmean$ and $\MSEspec$ scores, in that the estimates for points other than D2 and T2 are visually very close to the truth, while for D2 and T2 some differences are visible.

There are several explanations for the worse performance for D2 and T2 relative to the other points. One reason is that the spatial model has relatively little information to identify the existence of a cluster: these points belong to the category $\componentindex = 2$, which is represented by only 8 time series (in contrast to $\componentindex = 1, 3$ and $4$, which have 41, 18, and 33 members, respectively), and which has the smallest spatial area in the study. Another reason is that the process mean for $\componentindex = 2$ (equal to $1$ in the first half and $-1$ in the second half) is similar to that of the surrounding category, $\componentindex = 4$ (constant mean of $1$), making it harder to distinguish between the categories. For both reasons, it is not surprising to see worse performance for T2 and D2, which corresponds to good model behavior in the sense that it represents genuine model uncertainty. This is seen in Figure~\ref{fig:multiple_simulation_study_tvm}, where the $\hat{\mu}(\timeindex, \uvec)$ for T2 and D2 in the second half of the time series is shrunk towards that of $\componentindex = 4$, the enclosing category. It is also worth noting that the median $\MSEmean$ and $\MSEspec$ for T2 and D2 are small relative to the scales of their mean and log spectrum, respectively. This can also be seen qualitatively by the similarity between the true and estimated mean and spectra in figures~\ref{fig:multiple_simulation_study_tvm} and \ref{fig:multiple_simulation_study_tvs}.

\section{Applications}
\label{sec:applications}

In this section, we describe two applications. Section~\ref{sec:applications_rainfall} considers Australian rainfall, while incidence counts of measles in the United States are analyzed in Section~\ref{sec:applications_measles}.

\subsection{Australian rainfall data}
\label{sec:applications_rainfall}

Rainfall is governed in large part by cyclical processes, in particular the seasonal cycle driven by the Earth's orbit around the sun. It has therefore historically been a natural application area for spectral methods. These have been used to study interannual variation \citep[see, among many others,][]{alter1924,rajagopalanlall1998,anselletal2000}, intraannual or intraseasonal variation \citep{joshipandey2011}, and the connections between rainfall and other climatic processes \citep{rajagopalanlall1998,anselletal2000}. Here we focus on identifying changes in both the mean and spectrum of Australian rainfall. As part of a report on climate change tendered by several Australian government agencies, \citet{caietal2007} found that, since 1950, the Australian north has seen increased annual rainfall, while the southeast and southwest have experienced the opposite. The causes of these trends have been the subject of study and debate \citep[see, among others,][]{hopeetal2006,ummenhoferetal2009,pooketal2012,risbeyetal2013}. Apart from trends in overall rainfall, several authors have reported relative increases in heavy rainfall events, indicating changes in the variability of rainfall \citep{caietal2007,gallantetal2013}. We contribute to this literature by using AdaptSPEC-X to analyze the time varying mean and spectrum of Australian rainfall from sites dispersed over a wide spatial field, addressing simultaneously the question of whether changes have occurred in rainfall levels and rainfall variability.

\begin{figure}[ht!]
  \centering
  \begin{subfigure}[t]{\linewidth}
    \begin{minipage}[c]{0.10\textwidth}
      \caption{}
      \label{fig:monthly_rainfall_map}
    \end{minipage}
    \begin{minipage}[c]{0.825\textwidth}
      \centering
      \includegraphics{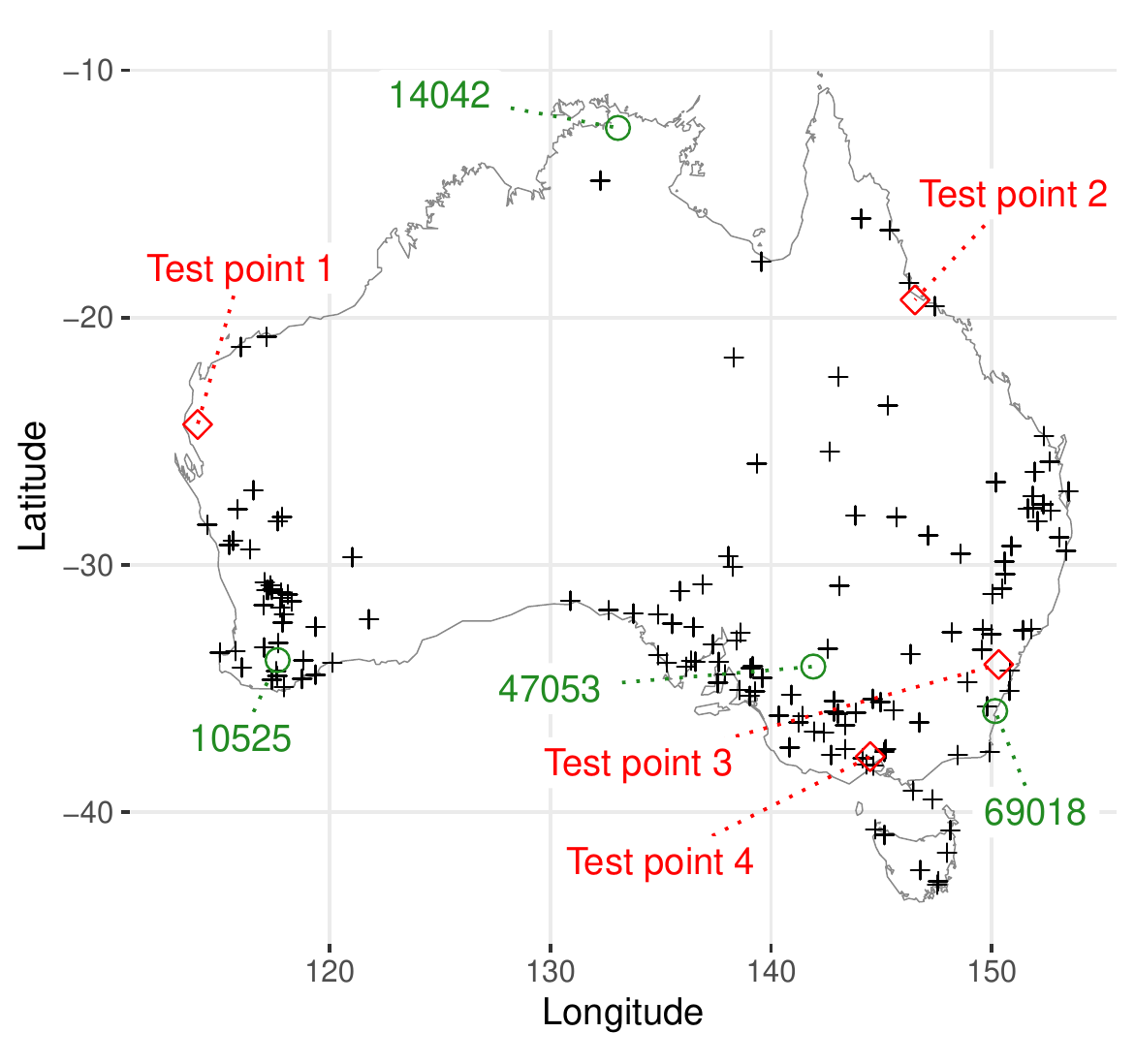}
    \end{minipage}\hfill
  \end{subfigure}

  \begin{subfigure}[t]{\linewidth}
    \begin{minipage}[c]{0.10\textwidth}
      \caption{}
      \label{fig:monthly_rainfall_ts}
    \end{minipage}
    \begin{minipage}[c]{0.90\textwidth}
      \centering
      \includegraphics{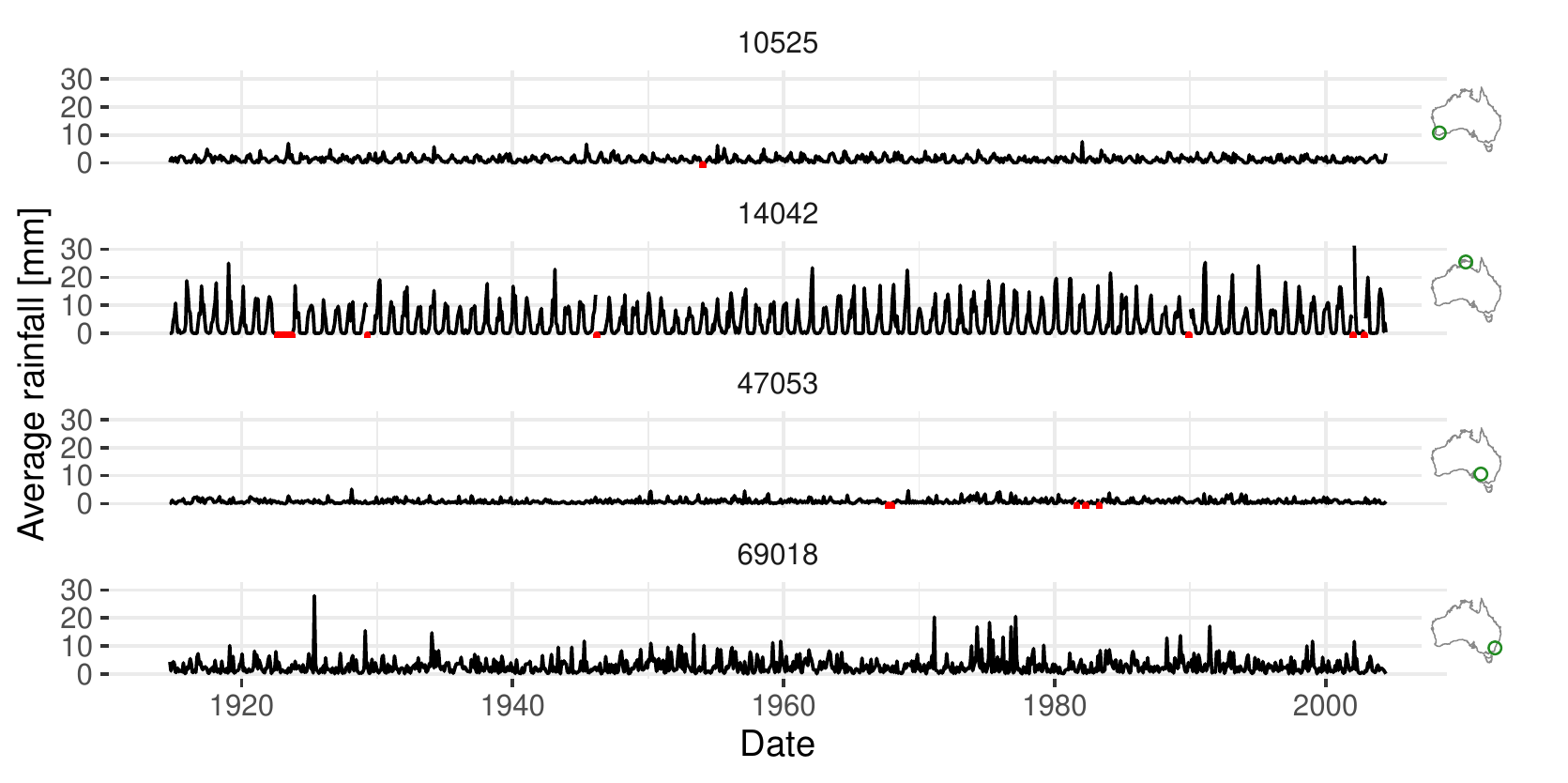}
    \end{minipage}\hfill
  \end{subfigure}

  \caption{
    (a) Locations of the 151 rainfall sites. Four example sites are marked with green circles, and four test locations are marked with red diamonds. (b) Monthly rainfall records at the four example sites, whose locations are indicated by inset maps. Missing values are marked with red ticks on the bottom axis.
  }
\end{figure}

We use data from \citet{bertolaccietal2019}, who studied the climatology of Australian daily rainfall using measurements from 17,606 sites across the continent. In particular, we use 151 of these sites characterized by having long and nearly contiguous rainfall records, the locations of which are displayed in Figure~\ref{fig:monthly_rainfall_map}.
These sites are among those identified by \citet{laveryetal1992} as having high quality records suitable for monitoring and assessing climate change. The raw time series are daily, and observations are typically made at 9 am local time, recording the total rainfall in millimeters (mm) for the previous 24 hours. Aggregation to monthly data is performed by calculating the average daily rainfall for the month.
To avoid artifacts, we consider as missing any month with fewer than fifteen days of measurements available (that is, not missing).

The resulting time series span the 1,078 months from September, 1914 to June, 2004 (inclusive), for a total of 162,778 observations, of which 4,095 are missing. The smallest possible measurement is 0 mm, corresponding to no rainfall for the month; this is true for 9,933 months. Time series for four example sites are displayed in Figure~\ref{fig:monthly_rainfall_ts}, and their locations are marked on inset maps (also marked in green in Figure~\ref{fig:monthly_rainfall_map}). The four time series span a wide range of average rainfall levels from around 1mm at site 47053 to 4mm at site 14042. They also exhibit varying levels of seasonality, where sites 10525 and 14042 have highly seasonal rainfall, while rainfall at sites 47053 and 69018 is less seasonal.

We fit AdaptSPEC-X to these data, setting $\uvec_\siteindex = (\text{lon}_\siteindex, \text{lat}_\siteindex)$, the longitude and latitude of the sites. Each mixture component has $\tmin = 60$ (5 years), chosen to ensure several observations of the dominant annual cycle are available. This constrains the maximum number of segments to be $M_\mathrm{max} = 17$. For the number of basis functions for the smoothing spline prior on the log spectra, we choose $J = 60$, giving sufficient flexibility to represent the spike at the dominant annual frequency. The support of the prior on $\mu^\componentindex_{s, m}$ is set to $(\mu_-, \mu_+) = (0, 30)$. The lower bound reflects the positivity of rainfall, while the upper bound is three times larger than the largest empirical mean attained in any 5-year period in the data. The LSBP is truncated at $\ncomponents = 25$ components, which was found to be large enough that higher values made no difference. Finally, the thin-plate GP prior for the LSBP has $B = 20$ basis functions, which captures more than 95\% of the variation implied by the prior. In addition to the locations with measurements, we estimate the predictive time varying mean and spectrum at four locations on the Australian landmass without measurements. These locations are indicated by red diamonds in Figure~\ref{fig:monthly_rainfall_map}.

Two major droughts occurred during the study period: the World War II drought of 1937--1945, and the Millennium drought that started in 1996 and was still ongoing by the end of the study period in 2004 \citep{ummenhoferetal2009}. Table~\ref{tab:monthly_rainfall_tests} presents estimated posterior probabilities of changes in the mean $\mu(\timeindex, \uvec)$ or variance $\sigma^2(\timeindex, \uvec) = 2 \int_0^{1/2} f(\timeindex, \omega, \uvec) d\omega $ around these times. Specifically, it shows that $\hat{P}(\mu_{1940} < \mu_{1950}) > 0.9$ and $\hat{P}(\sigma^2_{1940} < \sigma^2_{1950}) > 0.9$ at all four sites. However, $\hat{P}(\mu_{1940} < \mu_{1930}) < 0.7$ and $\hat{P}(\sigma^2_{1940} < \sigma^2_{1930}) < 0.57$ except for  site 69018 for which these probabilities are greater than 0.8. The Millennium drought is associated with drops in $\mu(\timeindex, \uvec)$, $\sigma^2(\timeindex, \uvec)$ or both at sites 14042, 47053 and 69018 as can be seen from the estimated probabilities at these sites: $\hat{P}(\mu_{2004} < \mu_{1990}) > 0.9$ and $\hat{P}(\sigma^2_{2004} < \sigma^2_{1990}) > 0.93$. Site 10525 in the southwest of the continent does not exhibit a drop with probability greater than $0.9$, consistent with the fact that the drought principally affected southeastern Australia \citep{ummenhoferetal2009}.

\citet{caietal2007} report large trends in rainfall since 1950. For southeast Australia, they report reductions in annual rainfall corresponding to 10--15mm/decade for the site 47053 and 50mm/decade for 69018. Consistent with this, Table~\ref{tab:monthly_rainfall_tests} shows that, for these sites, AdaptSPEC-X estimates that both $\mu(\timeindex, \uvec)$ and $\sigma^2(\timeindex, \uvec)$ declined between January, 1950 and January, 2004 with probability greater than $0.94$. The estimated drop in $\mu(\timeindex, \uvec)$ for site 47053 corresponds to a reduction of 1--16mm/decade\footnote{Calculated as $10 \times 365.25 \times \text{difference in daily average}\ /\ (2004 - 1950)$} (10th--90th percentile), consistent with \citeauthor{caietal2007}'s estimate. However, for site 69018, the estimated reduction is 6--42mm/decade, substantially less than 50mm/decade. For the site 14042 in the tropical north, \citeauthor{caietal2007} estimate increases of around 40mm/decade since 1950, while the estimates of Table~\ref{tab:monthly_rainfall_tests} indicate a decline over the same period. Finally, \citeauthor{caietal2007} report a decline of 20--30mm/decade in the region of southwest Australia containing the site 10525, but Table~\ref{tab:monthly_rainfall_tests} does not indicate a significant change in $\mu(\timeindex, \uvec)$ at this location (though the variance does appear to have declined). In contrast to \citeauthor{caietal2007} and \citet{gallantetal2013}, the reduction in $\sigma^2(\timeindex, \uvec)$ since 1950 at all locations suggests that rainfall variability has declined. This could have resulted from the use of different definitions of variability, e.g., counts of extreme events as in \citet{caietal2007}, versus our use of change in variance.

\begin{figure}
  \centering
  \includegraphics{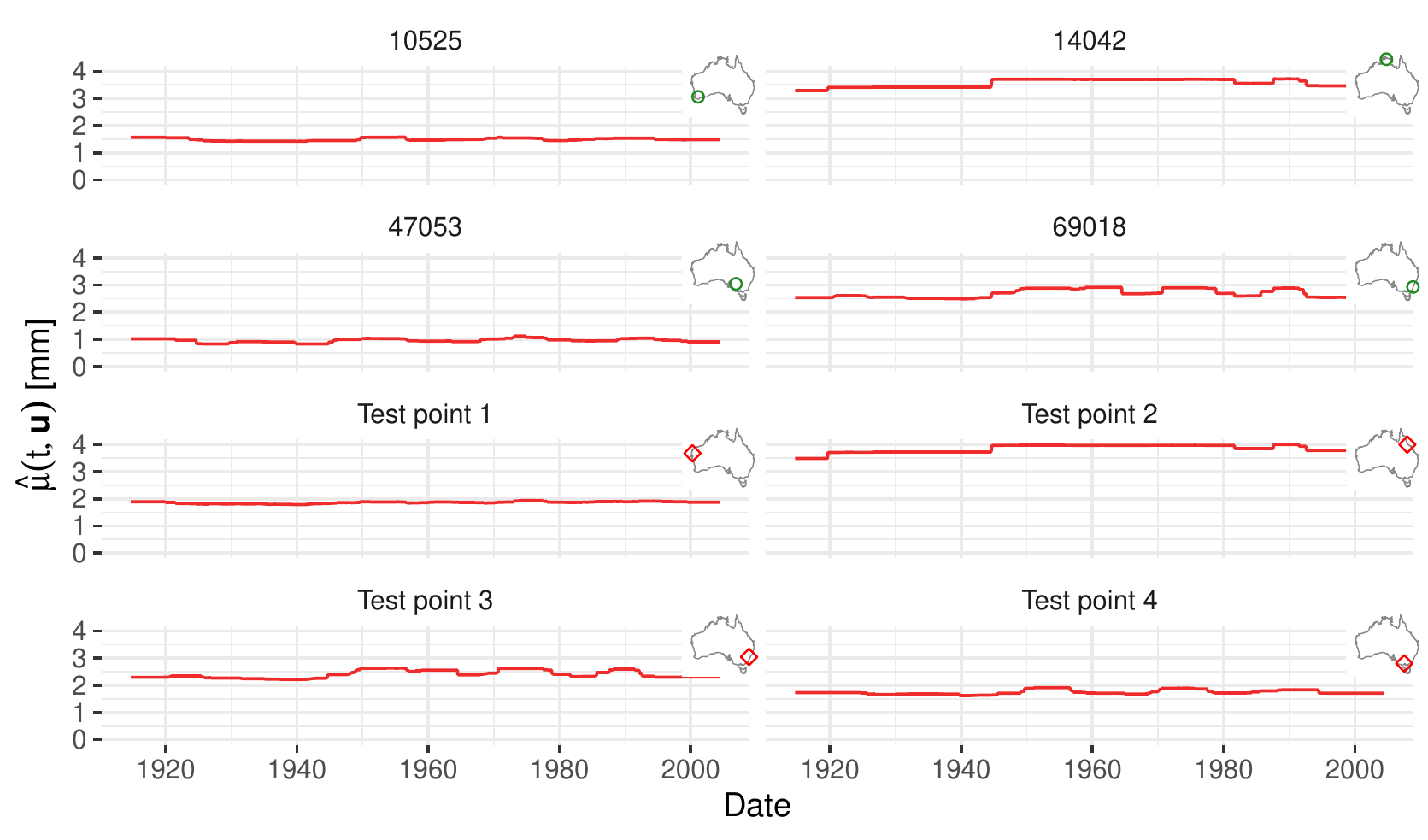}

  \caption{
    Estimated time varying means $\hat{\mu}(\timeindex, \uvec)$ for four monthly rainfall sites (first two rows), and four locations without observations (last two rows).
  }
  \label{fig:monthly_rainfall_tvm}
\end{figure}

\begin{figure}
  \centering
  \includegraphics{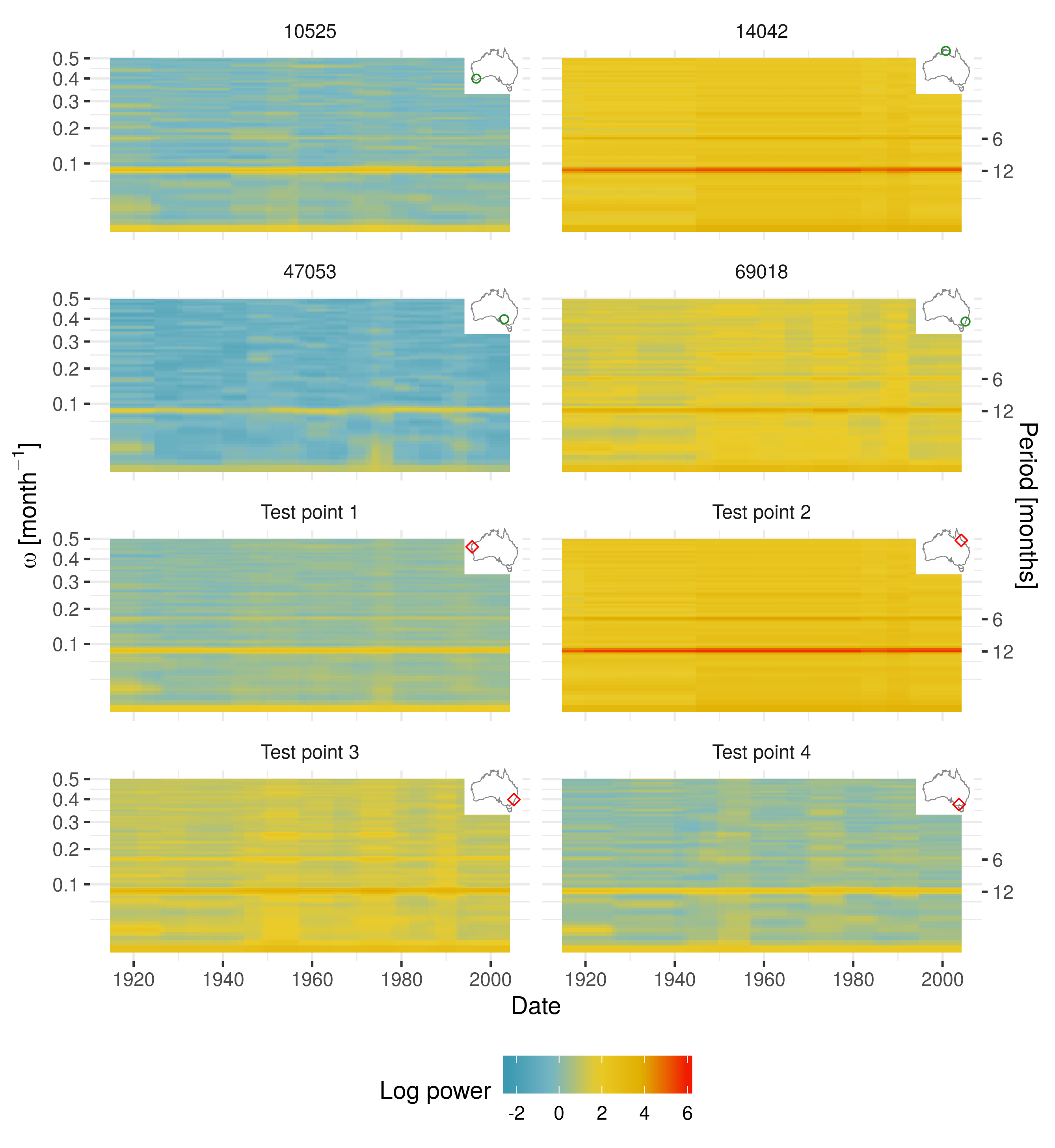}

  \caption{
    Estimated time varying spectra $\log \hat{f}(\timeindex, \omega, \uvec)$ for four monthly rainfall sites (first two rows) and four locations without observations (last two rows). The color indicates the log power at the corresponding time and frequency. The $\omega$-axis is on a square-root scale. The axis on the right-hand side displays the period $(1 / \omega)$.
  }
  \label{fig:monthly_rainfall_tvs}
\end{figure}

\begin{figure}
  \centering
  \includegraphics{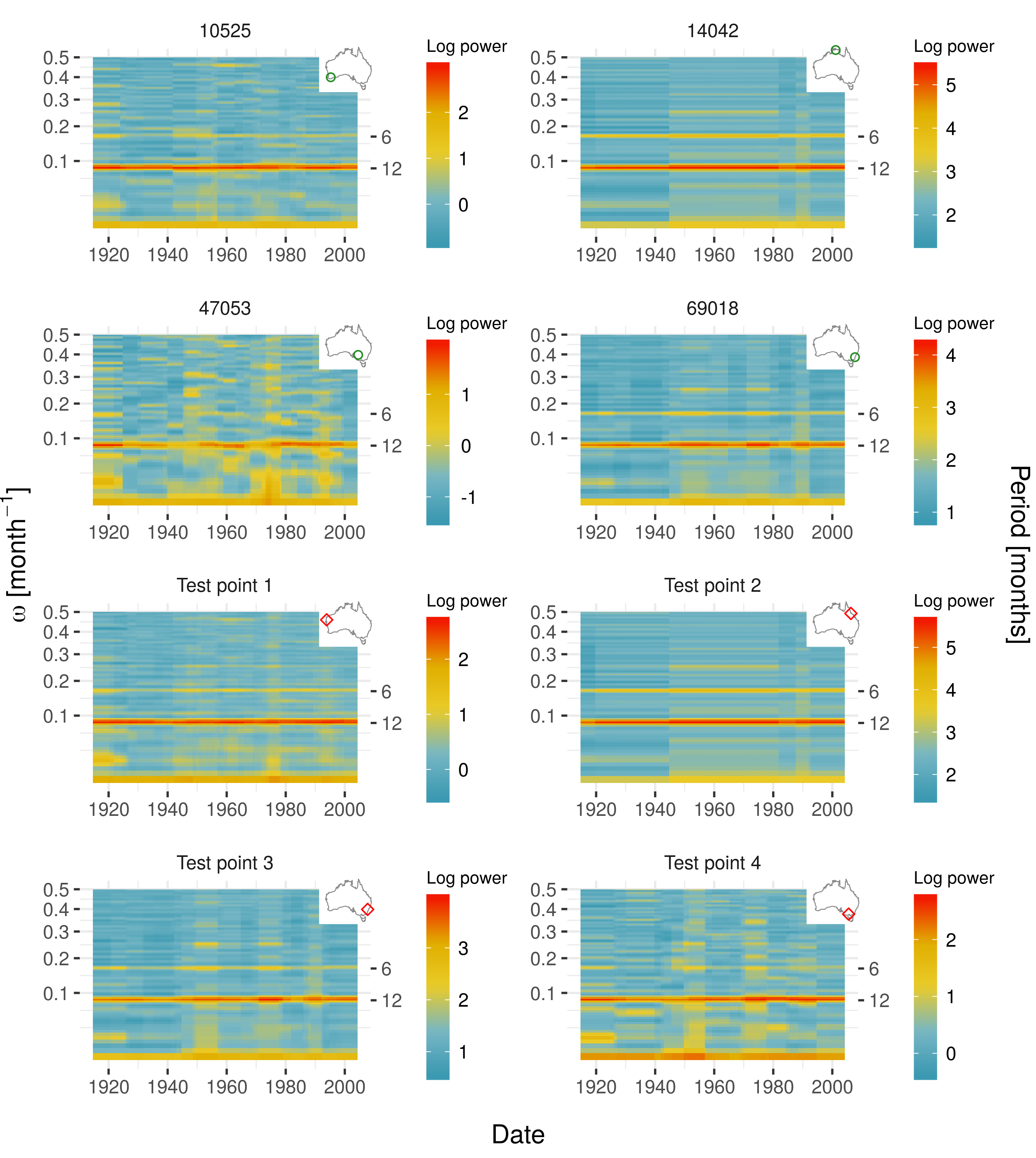}

  \caption{
    The same estimated time varying spectra $\log \hat{f}(\timeindex, \omega, \uvec)$ as in Figure~\ref{fig:monthly_rainfall_tvs}, except that each location is given its own color scale.
  }
  \label{fig:monthly_rainfall_tvs_unscaled}
\end{figure}

\begin{table}
  \centering
  \bgroup
\def\arraystretch{1.2}
\begin{tabular}{ll|rrrr}
& & \multicolumn{4}{c}{Site ($\uvec$)} \\
Event & $\hat{p}(\cdot \mid \xvec)$ & 10525 & 14042 & 47053 & 69018 \\ \hline 
\hline 
 \multirow{4}{*}{WW2 drought}
& $\mu(\text{1940-01}, \uvec) < \mu(\text{1930-01}, \uvec)$ & 0.539 & 0.442 & 0.700 & 0.802 \\
& $\sigma^2(\text{1940-01}, \uvec) < \sigma^2(\text{1930-01}, \uvec)$ & 0.572 & 0.390 & 0.508 & 0.833 \\
& $\mu(\text{1940-01}, \uvec) < \mu(\text{1950-01}, \uvec)$ & 0.915 & 0.980 & 0.997 & 0.994 \\
& $\sigma^2(\text{1940-01}, \uvec) < \sigma^2(\text{1950-01}, \uvec)$ & 0.997 & 0.996 & 1.000 & 0.997 \\
\hline 
 \multirow{2}{*}{Millenium drought}
& $\mu(\text{2004-01}, \uvec) < \mu(\text{1990-01}, \uvec)$ & 0.878 & 0.906 & 0.975 & 0.942 \\
& $\sigma^2(\text{2004-01}, \uvec) < \sigma^2(\text{1990-01}, \uvec)$ & 0.747 & 0.932 & 0.999 & 0.978 \\
\hline 
 \multirow{2}{*}{Long term}
& $\mu(\text{2004-01}, \uvec) < \mu(\text{1950-01}, \uvec)$ & 0.818 & 0.907 & 0.940 & 0.972 \\
& $\sigma^2(\text{2004-01}, \uvec) < \sigma^2(\text{1950-01}, \uvec)$ & 0.961 & 0.954 & 0.999 & 0.989 \\
\end{tabular}
\egroup

  \caption{Estimated posterior probabilities $\hat{p}(\cdot \mid \xvec)$ of various events for monthly rainfall at the four example sites. The events correspond to the WW2 drought (first four rows), the Millenium drought (second two rows), and long term change (last two rows). For each event and site ($\uvec$), the table presents probabilities that the mean $\mu(\timeindex, \uvec)$ and the variance $\sigma^2(\timeindex, \uvec)$ changed before or after the event.}
  \label{tab:monthly_rainfall_tests}
\end{table}

\clearpage

\subsection{Measles incidence in the United States}
\label{sec:applications_measles}

Measles is a highly contagious disease that causes fever, cough, runny nose, and a rash \citep{moreno2018}. Complications of measles can include pneumonia, deafness, or death \citep{moreno2018}. Prior to the licensing of a vaccine in 1963, the incidence rate of measles averaged 318 cases per 100,000 population per year, with outbreaks occurring annually or every other year \citep{vanpanhuisetal2013}. After vaccine licensure, incidence declined dramatically, and endemic measles transmission was declared eliminated from the United States in 2000 \citep{katzhinman2004}. As part of a study on the impact of vaccination for a variety of contagious diseases, \citet{vanpanhuisetal2013} collated a unique data set by digitizing weekly surveillance reports from the United States of several nationally notifiable diseases, including measles, and have made these data available online at the Project Tycho website\footnote{\url{https://www.tycho.pitt.edu/}}. In this section we analyze Project Tycho's measles data using AdaptSPEC-X.

\begin{figure}
  \centering
  \includegraphics[angle=90]{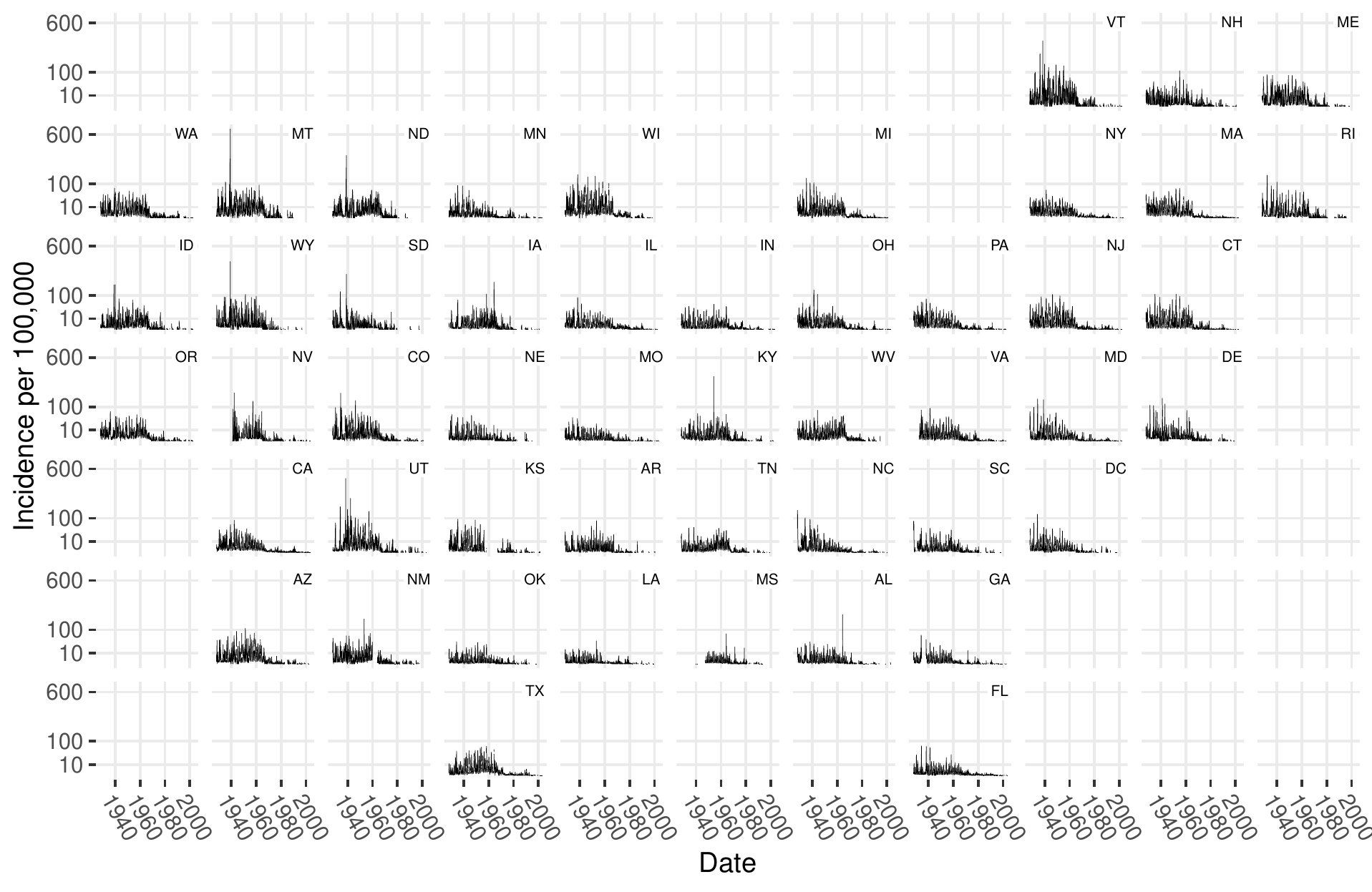}

  \caption{
    Measles incidence rate per 100,000 population for the continental US (that is, excluding Hawaii and Alaska), plus the District of Columbia. Each panel shows one state, where the layout of the panels roughly matches the geographic distribution of the states. The incidence axis is square-root transformed.
  }
  \label{fig:measles_ts}
\end{figure}

The data comprise weekly time series of measles incidence for each state, reporting the weekly (where the week starts on a Sunday) incidence rate per 100,000 population. In this work we use time series from the continental United States (that is, excluding Hawaii and Alaska), plus the District of Columbia. These span the 3,914 weeks from week one of 1928 (which we write as 1928-01) to week one of 2003 (2003-01). Across all 49 time series there are 191,786 observations. Of these, 50,067 (26\%) are missing, and 30,439 (16\%) have incidence equal to zero. The time series are shown in Figure~\ref{fig:measles_ts}, where each panel presents the series for a state, and
the incidence axis is on a square-root scale (note however that the spectral analysis performed later is applied to the untransformed data). The layout of the panels roughly matches the geographic distribution of the states. The most striking aspect of these plots is the dramatic decline in both the level and volatility of incidence starting in 1963, the year of vaccine licensure.

\begin{figure}[!ht]
  \centering
  \includegraphics[angle=90]{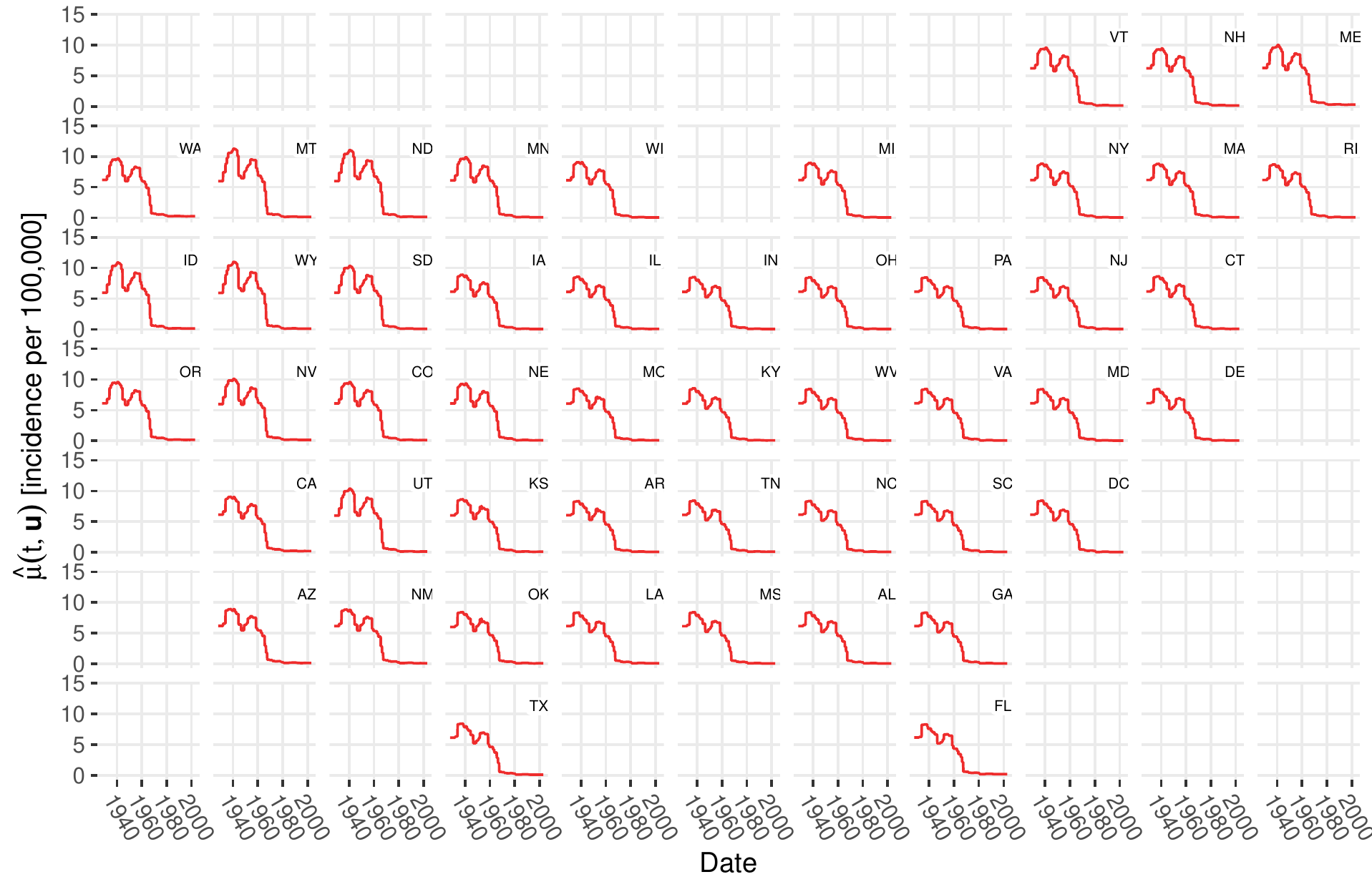}

  \caption{Estimated time varying mean $\hat{\mu}(\timeindex, \uvec)$ for measles incidence, where each panel shows the estimate for one state.}
  \label{fig:measles_tvm}
\end{figure}

\begin{figure}
  \centering
  \includegraphics[angle=90]{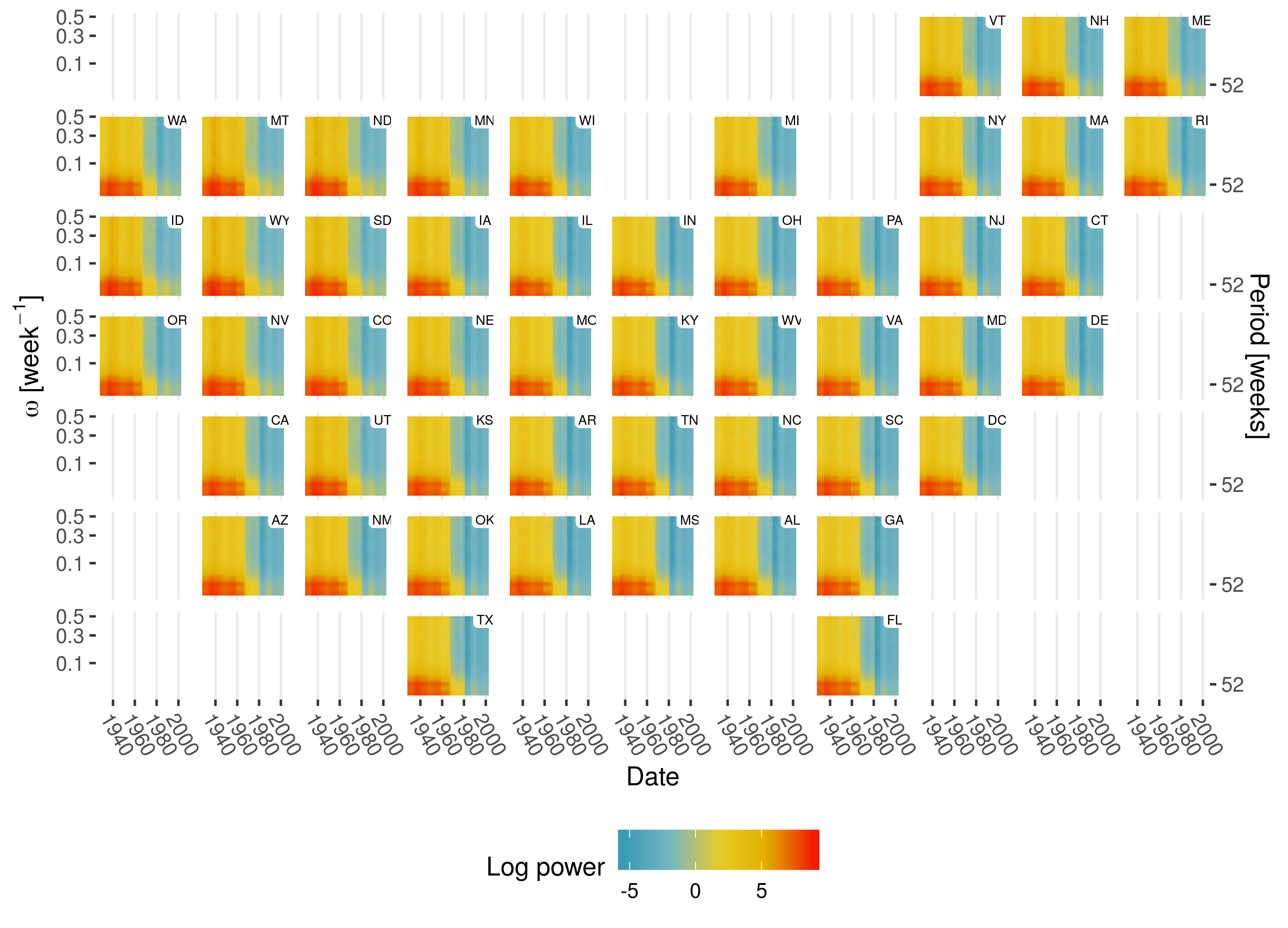}

  \caption{
    Estimated time varying log spectra, $\log \hat{f}(\timeindex, \omega, \uvec)$ for measles incidence, where each panel shows the estimate for one state. Colors indicate the log power at the corresponding date and $\omega$. The $\omega$-axis is on a square-root scale. The top axis displays the period ($1 /\omega$).
  }
  \label{fig:measles_tvs}
\end{figure}

\begin{figure}
  \centering
  \includegraphics{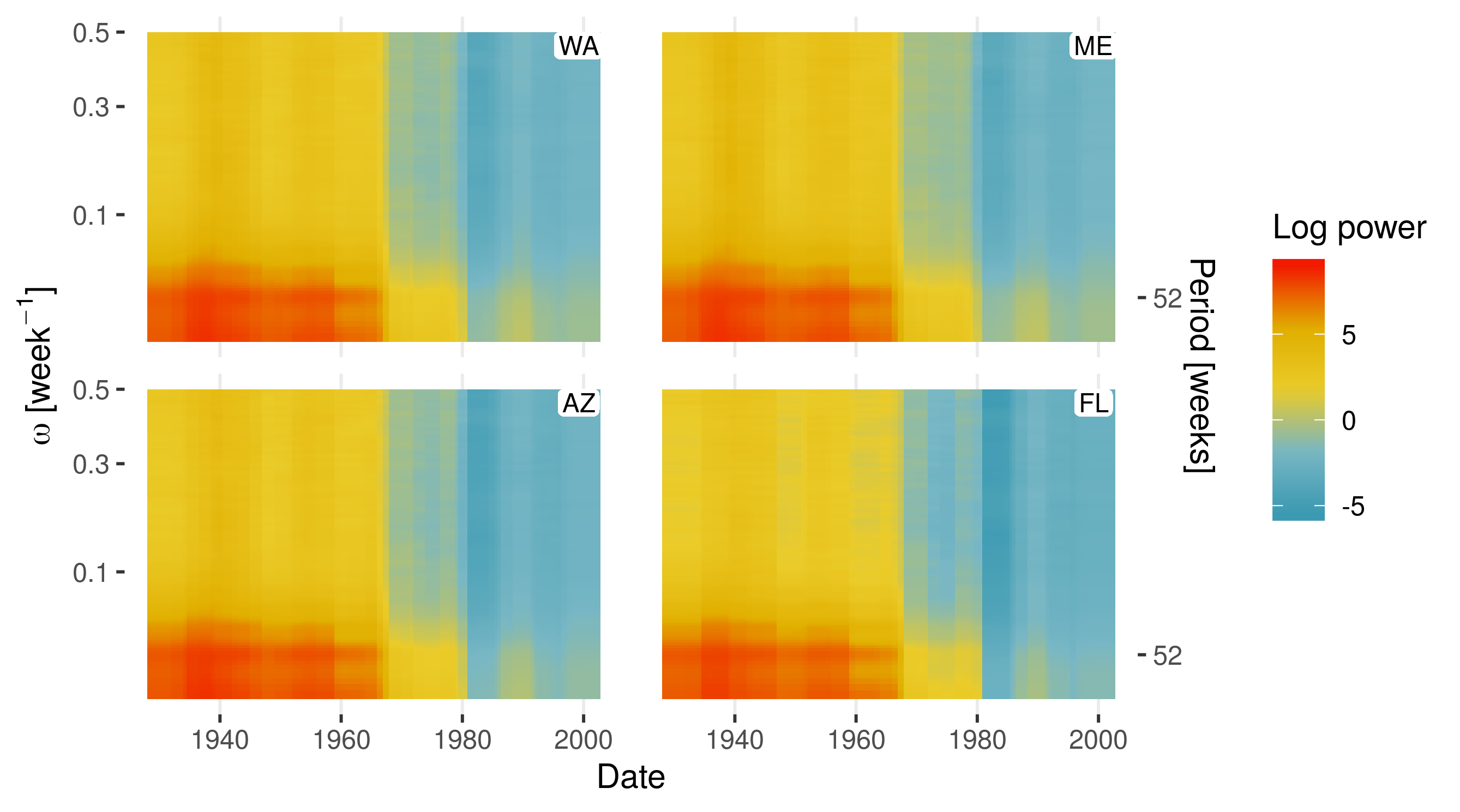}
  \includegraphics{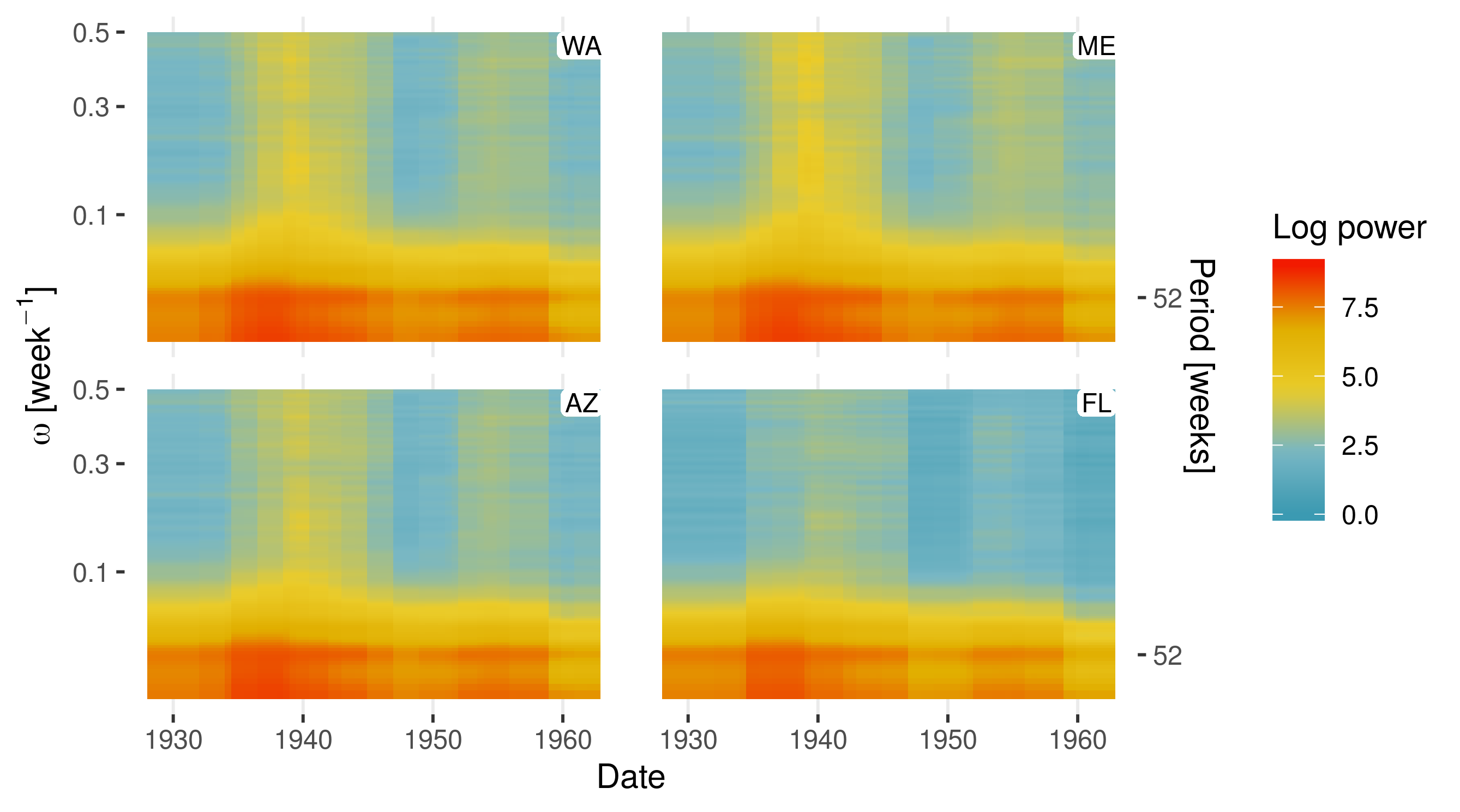}

  \caption{
    Estimated time varying log spectra, $\log \hat{f}(\timeindex, \omega, \uvec)$, for four states. Estimates for the full study period (as in Figure~\ref{fig:measles_tvs}) are shown in the top two rows, while the bottom two rows display estimates for the pre-vaccine period (as in Figure~\ref{fig:measles_tvs_before} below).
  }
  \label{fig:measles_tvs_special}
\end{figure}

We set $\uvec_\siteindex = (\text{lon}_\siteindex, \text{lat}_\siteindex)$, the longitude and latitude of the centroid of each state, and fit AdaptSPEC-X to the measles time series. Each mixture component has $\tmin = 208$ (4 years) as the minimum segment length. This was chosen to ensure four observations of the annual cycle in each segment. The maximum number of segments was set as $M_\mathrm{max} = 18$, the maximum allowed by the combination of $\tmin$ and the number of weeks in the data. The support of the prior on $\mu^\componentindex_{s, m}$ was set to $(\mu_-, \mu_+) = (0, 20)$; the lower bound represents the known positivity of the incidence rates (and is particularly helpful to constrain $\mu^\componentindex_{s, m}$ in periods with very small counts), while the upper bound is twice as large as the posterior mean of this parameter. As in the rainfall application, the number of basis functions for the smoothing spline prior on the log spectra was set to $J = 60$. This was found to be high enough to represent the spectrum accurately, particularly the spike corresponding to the annual cycle. The LSBP is truncated at $\ncomponents = 10$ components, as we found higher values did not change the results. Finally, the thin-plate GP prior for the LSBP has $B = 20$ basis functions, which captures more than 95\% of the variation implied by the prior. The resulting estimated time varying means and spectra are shown in figures~\ref{fig:measles_tvm} and \ref{fig:measles_tvs}, respectively. The top two rows of Figure~\ref{fig:measles_tvs_special} display zoomed-in plots of the time varying spectra for four states: Arizona, Florida, Maine, and Washington.

As in Figure~\ref{fig:measles_ts}, the post-vaccine drop in the mean and power of incidence is the most obvious feature. This drop occurs in steps, starting with a dramatic drop following the licensing of the Edmonston vaccine in 1963, stalling around 1970, then dropping again around 1980. This corresponds to the waxing and waning of government funding and effort targeted at measles elimination, which culminated in an intensified elimination drive \citep{hinmanetal1979,atkinsonetal1992}. An outbreak during the early 1990s is visible as an increase in power in Figure~\ref{fig:measles_tvs}; this outbreak received much attention and resulted in changes to the immunization schedule for children \citep{atkinsonetal1992}.

In the pre-vaccine period, the spectra in Figure~\ref{fig:measles_tvs} have peaks around frequencies $1 / 52$ and 0, indicating annual seasonality and long-term dependence, respectively. After the introduction of the vaccine, the annual peak disappears. \citet{grenfelletal2001} identified a biennial cycle in similar measles data for the UK, but this does not appear to be a feature of the US data. Figures~\ref{fig:measles_tvm} and \ref{fig:measles_tvs} also indicate changes in the mean and spectrum during the pre-vaccine years, where all states exhibit periods of increased mean incidence and power centered around 1940 and 1955. This concords with \citet{vanpanhuisetal2013}, who noted that incidence rates had variable patterns in the pre-vaccine time period, speculating that these may have been due to sanitation, hygiene, or demographic factors.

\begin{figure}
  \centering
  \includegraphics[angle=90]{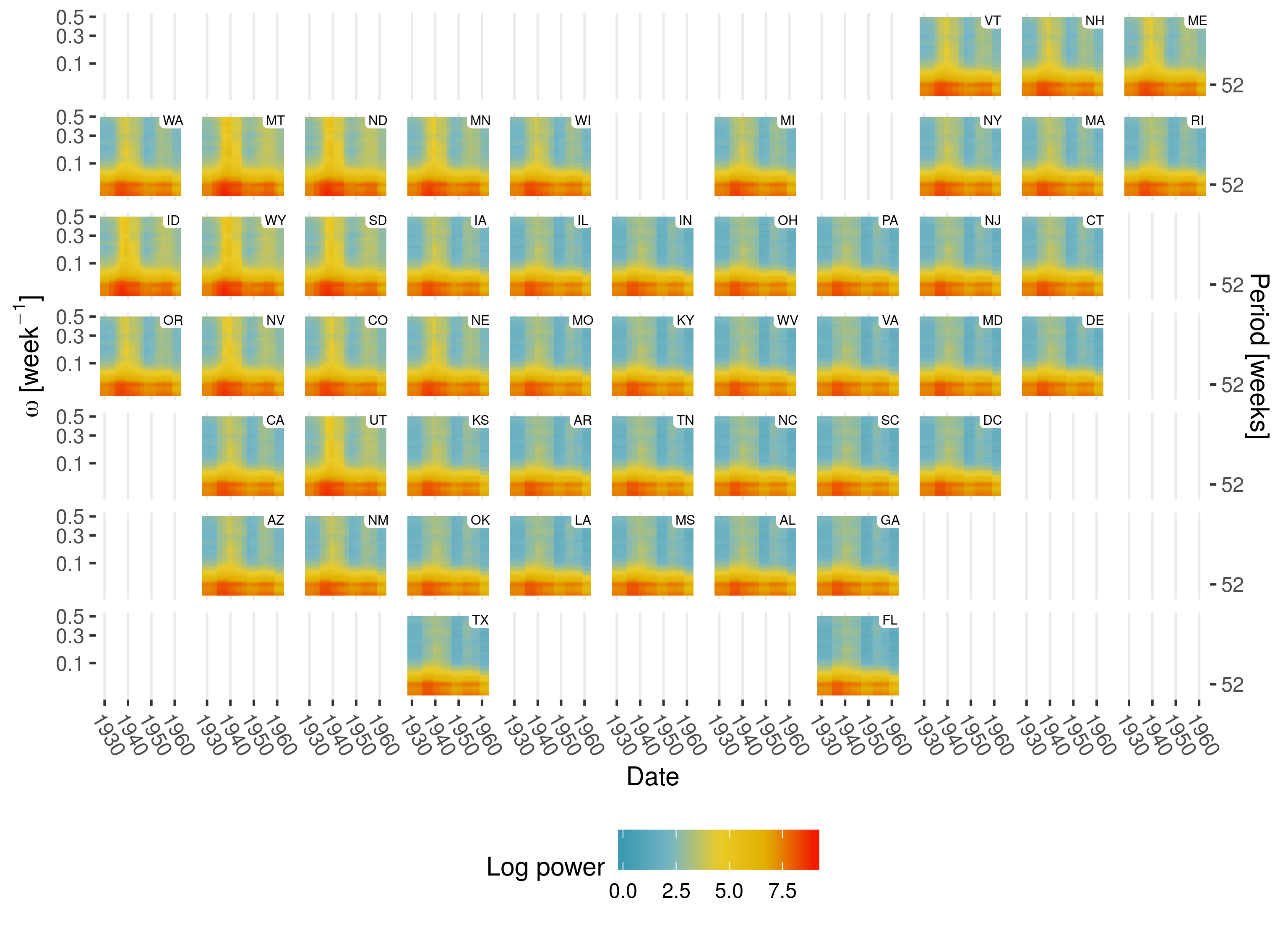}

  \caption{
    Estimated time varying log spectra, $\log \hat{f}(\timeindex, \omega, \uvec)$ for measles incidence for the pre-vaccine period ($<$1963). Figure~\ref{fig:measles_tvs} shows the estimates for the full study period.
  }
  \label{fig:measles_tvs_before}
\end{figure}

Because of the extreme nonstationarity introduced by the vaccine, the time varying spectra in Figure~\ref{fig:measles_tvs} span such a wide range of powers that the spectra for all states look almost identical. This is not the case, as shown in Figure~\ref{fig:measles_tvs_before} and in the bottom two rows of Figure~\ref{fig:measles_tvs_special}, which display time varying spectra for the pre-vaccine years only. These spectra highlight the existence of geographic heterogeneity between states, where higher power is more typical of the west and north, compared to the south and east.

Since the elimination of endemic measles in the US in 2000, there have been a number of outbreaks associated with individuals `importing' measles by acquiring the disease while outside the US and spreading it upon their return \citep{parkeretal2006,cdc2019}. \citet{phadkeetal2016} associated several of these outbreaks with individuals unvaccinated for nonmedical reasons, which they term as vaccine refusal. Some authors have even declared that a resurgence of measles has occurred \citep{lynfielddaum2014}. Using the AdaptSPEC-X fits, we tested whether the mean $\mu(\timeindex, \uvec)$ or variance $\sigma^2(\timeindex, \uvec)$ of measles increased from 1995-01, a few years after the big outbreak in the early 1990s, to 2003-01, the last time period in the data. We find no evidence of increase in $\mu(\timeindex, \uvec)$ in any state, which can be seen from the fact that the highest posterior probability of an increase equals 0.68 in Oklahoma. As for $\sigma^2(\timeindex, \uvec)$, the posterior probabilities of an increase in all states range between 0.81--0.87, which we consider to be weak evidence of change. Unfortunately, because the data end in 2003, it is not possible to assess changes to the mean or spectrum of measles incidence in more recent years.

\clearpage

\section{Discussion}

This article has presented AdaptSPEC-X, a Bayesian method for analyzing a panel of possibly nonstationary time series using a covariate-dependent infinite mixture model, with mixture components parameterized by their time varying mean and spectrum. AdaptSPEC-X extends AdaptSPEC to accommodate multiple time series, each with its own covariate values. Specifically, the covariates, which are assumed to be time-independent, are incorporated via the mixture weights using the logistic stick breaking process. The mixture components are based on AdaptSPEC, which handles a single nonstationary time series. In particular, it partitions a time series into an unknown but finite number of segments, and estimates the spectral density within each segment by smoothing splines. New features which have been added to the AdaptSPEC components include estimation of time varying means and handling of missing observations. The model and sampling scheme can accommodate large panels, such as that of the measles application. In addition to estimating time varying spectra for each time series in the panel, AdaptSPEC-X allows inference about the underlying process at unobserved covariate values, enabling predictive inference. Efficient software implementing AdaptSPEC-X is available in the R package BayesSpec.

In Section~\ref{sec:model_log_odds}, the log odds of the LSBP, which determine the mixture weights, are modeled using a thin-plate GP prior. While this prior is flexible, it is also smooth and stationary. This property may be inappropriate in settings where changes in the mean or spectrum of the individual time series occur abruptly over the covariate space. An extension to a nonstationary prior for the log odds, or a piecewise prior as in \citet{bruceetal2018}, may be of interest in these cases.

AdaptSPEC (and therefore AdaptSPEC-X) relies on Whittle's approximation to the likelihood (Equation~\eqref{eqn:whittle_likelihood}). The Whittle likelihood is asymptotically correct for both Gaussian and non-Gaussian time series \citep{hannan1973}, but is known to be inefficient for small sample sizes \citep{contrerasetal2006}. Several methods exist in the literature to ameliorate this problem \citep[see, for example,][]{sykulskietal2019}, and these methods might produce useful extensions to AdaptSPEC for settings with short time series or small segment lengths (i.e., small $\tmin$).

Neither AdaptSPEC nor AdaptSPEC-X account explicitly for measurement error. For i.i.d.\@ Gaussian measurement error, this should not cause a problem, as the added variance would appear as an added constant (that is, white noise) in the spectrum. On the other hand, if the measurement error is changing over time, this may be estimated as false nonstationary, in the sense that the underlying process is not actually changing. Adding an explicit layer for measurement error to AdaptSPEC-X would be a useful extension for these cases. More specialized forms of measurement error may be useful in other settings. For example, in the measles application of Section~\ref{sec:applications_measles}, the data are incidence rates per 100,000 population. While rates are continuous quantities, they are constructed from counts, so the measurement process is actually discrete. A hierarchical extension accounting for this would be an interesting addition to AdaptSPEC-X.

AdaptSPEC-X allows for covariates that do not vary with time, and an extension to a more general framework accommodating time varying covariates would facilitate new types of inference. For example, external time varying climate indicators such as the Southern Oscillation Index are known to influence rainfall patterns \citep{bertolaccietal2019}. A time varying covariate influencing the mean or the spectrum might introduce useful shrinkage, improving predictive performance. One challenge with such an approach would be to determine how a time varying covariate could interact with the segmentation approach used by AdaptSPEC to handle nonstationarity. It would also be interesting to allow the mean to be time varying and covariate-dependent within segments, not only between segments. Future research will focus on these extensions.

\ifdefined\isblinded

\else
  \section*{Reproducibility}

  Data and code reproducing the figures and tables in this manuscript are available online at \url{https://github.com/mbertolacci/adaptspecx}.

  \section*{Funding}

  E. Cripps and M. Bertolacci were supported by the ARC Industrial Transformation Research Hub for Offshore Floating Facilities which is funded by the Australian Research Council, Woodside Energy, Shell, Bureau Veritas and Lloyds Register (Grant No. 140100012). S. Cripps is the recipient of an Australian Research Council Australian Future Fellowship (140101266) funded by the Australian Government. O. Rosen was supported in part by grants NSF DMS-1512188 and NIH 2R01GM113243-05.
\fi

\bibliographystyle{asa}
\bibliography{adaptspecx_paper}{}

\appendix
\renewcommand{\theequation}{A.\arabic{equation}}
\setcounter{equation}{0}

\section{Conditional distributions used for the MCMC scheme}
\label{sec:conditional_distributions}

This appendix provides the details of the conditional distributions sampled in the MCMC scheme of Section~\ref{sec:sampling_scheme}, with the exception of the distribution of $(\xvec^\all_\mis \mid \betavec^\dagger, \Theta, \zvec, \xvec^\all_\obs)$, which is described in Section~\ref{sec:model_missing_values}.

\subsection{\texorpdfstring{Drawing $\Theta_\componentindex$}{Theta}}
\label{sec:conditional_distribution_theta}

The conditional distribution is
\[
  p(\Theta_\componentindex \mid \zvec, \xvec^\all) \propto p(\Theta_\componentindex) \prod_{\{\siteindex:\ z_\siteindex = \componentindex\}} g_\componentindex(\xvec_\siteindex \mid \Theta_\componentindex).
\]
This step is potentially transdimensional as the number of segments $m$ may increase or decrease, so that $\Theta_\componentindex$ may change dimension. RWS12 describe a reversible-jump transition kernel for $\Theta_\componentindex$; the interested reader can find the details therein.

The next two sections describe modifications to the transition kernel of RWS12. The first modification samples the segment means, $\muvec^\componentindex_m$, and is described in Section~\ref{sec:sampling_scheme_means}. The second modification is the incorporation of an RMHMC step to accelerate convergence, described in Section~\ref{sec:sampling_scheme_rmhmc}. Some additional details from RWS12 are needed. For ease of exposition, in what follows we return to the case of $\xvec$ stationary, omitting the component indicator superscript $\componentindex$ and the segment subscript $s, m$. RWS12 express the smoothing spline prior on $\log f(\omega)$ (see Section~\ref{sec:model_single}) using the linear basis expansion,
\[
  \log f(\omega_\freqindex) = \alpha_0 + h(\omega_\freqindex) = \qvec_\freqindex' \bvec,
\]
where
\[
  \qvec_\freqindex = \left(
    1,
    \frac{\sqrt{2} \cos(2\pi \omega_\freqindex)}{\pi},
    \frac{\sqrt{2} \cos(4\pi \omega_\freqindex)}{2\pi},
    \ldots,
    \frac{\sqrt{2} \cos(2 J \pi \omega_\freqindex)}{J\pi}
  \right)'
\]
and $\bvec = (\alpha_0, b_1, \ldots, b_J)'$. The prior on $\bvec$ is $(\bvec \mid \tau_b^2) \sim N(\bm{0}, \Sigma_b)$, where
\[
  \Sigma_b = \begin{pmatrix}
    \sigma^2_\alpha & 0  \\
    0 & \tau_b^2 I_J
  \end{pmatrix},
\]
$\sigma^2_\alpha$ is a hyperparameter, and $\tau_b^2$ is a smoothing parameter (having its own prior, see RWS12). Note that the symmetry of the spectral density is such that $f(\omega_\freqindex) = f(\omega_{\ntimes - \freqindex + 2})$. Thus, later in this section when we write $\qvec_\freqindex$ for $\freqindex > \ntimes / 2 + 1$, we refer to $\qvec_{\ntimes - \freqindex + 2}$. In general, each segment $(s, m)$ of each component $\componentindex$ has an associated $\bvec^\componentindex_{s, m}$ and $\tau^{2\componentindex}_{b, s, m}$.

\subsubsection{\texorpdfstring{Drawing $\mu$}{Segment mu}}
\label{sec:sampling_scheme_means}

The conditional distribution of $\mu$ is
\begin{equation}
  \begin{split}
    p(\mu \mid \xvec, f)
      & \propto p(\xvec \mid f, \mu) p(\mu) \\
      & \propto
      \exp\left\{
        -\frac{1}{2} \sum_{\freqindex = 1}^\ntimes \frac{I_\freqindex}{f(\omega_\freqindex)}
      \right\} p(\mu) \\
      & \propto
      \exp\left\{ -\frac{1}{2} \frac{I_1}{f(\omega_1)} \right\} \ind(\mu_- < \mu < \mu_+) \\
      & \propto
      \exp\left\{ -\frac{1}{2} \frac{(\bar{x} - \mu) ^ 2}{f(\omega_1) / \ntimes} \right\} \ind(\mu_- < \mu < \mu_+),
  \end{split}
  \label{eqn:mu_conditional_derivation}
\end{equation}
where $\bar{x}$ is the arithmetic mean of $\xvec$, and $\ind(\cdot)$ is an indicator function. The third line of Equation~\eqref{eqn:mu_conditional_derivation} follows from the fact that only $I_1$ depends on $\mu$, and the final line follows from $I_1 = \ntimes(\bar{x} - \mu)^2$. Therefore,
\begin{equation}
  (\mu \mid \xvec, f) \sim N_{\mu_-, \mu_+}(\bar{x}, f(\omega_1) / \ntimes), \label{eqn:mu_conditional}
\end{equation}
where $N_{\mu_-, \mu_+}$ denotes the truncated normal distribution on $(\mu_-, \mu_+)$.

This distribution is incorporated in two places in the transition kernel of RWS12. First, the proposal distribution for $\bvec$ based on the mode of its full conditional used both for the between- and within- model steps (see Appendix of RWS12) is adjusted to propose from the mode of $(\mu, \bvec')'$, where the modal value for $\mu$ is $\bar{x}$. Second, an additional step is added to each iteration of the MCMC scheme, in which a segment $s \in \{ 1, \ldots, m \}$ is chosen uniformly at random, and the corresponding $\mu_{s, m}$ is updated based on Equation~\eqref{eqn:mu_conditional}. This latter step is simply to aid convergence of the resulting Markov chain, and is necessary to ensure ergodicity.

\subsubsection{\texorpdfstring{Drawing $\bvec$}{b} using RMHMC}
\label{sec:sampling_scheme_rmhmc}

\citet{girolami2011} propose a variant of the Hamiltonian Monte Carlo (HMC) sampling method \citep{duaneetal1987,neal1993} for sampling from an unnormalized probability density: Riemann manifold HMC (RMHMC). RMHMC exploits the Riemann geometry of the parameter space of the density in order to efficiently traverse the space and thereby provide an efficient proposal mechanism without the need for tuning. These authors propose the use of the Fisher information plus the negative Hessian of the log prior to define a metric tensor over the parameter space, and show how to modify plain HMC to incorporate this geometry. They show that the performance of RMHMC is excellent, comparable or better than that of competing methods, while requiring less tuning. In this section, we show how an RMHMC step for the smoothing spline parameters $\bvec$ may be incorporated into the sampling scheme of RWS12, improving the convergence of the scheme.

The appropriate metric tensor for the conditional distribution of $\bvec$ is
\[
  -E_{\xvec} \left[ \frac{\partial^2}{\partial \bvec \partial \bvec'} \log p(\bvec \mid \xvec, \mu, \tau^2_b) \right],
\]
where $E_{\xvec}$ means that the expectation is with respect to $\xvec$. We now derive this expectation analytically. The conditional distribution for $\bvec$ is
\begin{align*}
  p(\bvec \mid \xvec, \mu, \tau^2_b)
    \propto p(\xvec \mid \bvec, \mu)p(\bvec \mid \tau^2_b)
  & \propto
    \frac{1}{\prod_{\freqindex = 1}^\ntimes f(\omega_\freqindex)^{1 / 2}}
    \exp\left\{
      -\frac{1}{2} \sum_{\freqindex = 1}^\ntimes \frac{I_\freqindex}{f(\omega_\freqindex)}
      -\frac{1}{2} \bvec' \Sigma_b^{-1} \bvec
    \right\} \\
  & \propto
    \frac{1}{\prod_{\freqindex = 1}^\ntimes e^{\frac{1}{2} \qvec_\freqindex' \bvec}}
    \exp\left\{
      -\frac{1}{2} \sum_{\freqindex = 1}^\ntimes \frac{I_\freqindex}{e^{\qvec_\freqindex' \bvec}}
      -\frac{1}{2} \bvec' \Sigma_b^{-1} \bvec
    \right\}.
\end{align*}
The first and second derivatives of $\log p(\bvec \mid \xvec, \mu, \tau^2_b)$ are
\begin{equation}
  \begin{split}
    \frac{\partial}{\partial \bvec} \log p(\bvec \mid \xvec, \mu, \tau^2_b)
    & = \frac{1}{2}\sum_{\freqindex = 1}^\ntimes \qvec_\freqindex \left( \frac{I_\freqindex}{e^{\qvec_\freqindex' \bvec}} - 1 \right)
        - \Sigma_b^{-1} \bvec, \text{ and} \\
    \frac{\partial^2}{\partial \bvec \partial \bvec'} \log p(\bvec \mid \xvec, \mu, \tau^2_b)
    & = - \frac{1}{2} \sum_{\freqindex = 1}^\ntimes \frac{I_\freqindex}{e^{\qvec_\freqindex' \bvec}} \qvec_\freqindex' \qvec_\freqindex
        - \Sigma_b^{-1},
  \end{split}
  \label{eqn:b_derivatives}
\end{equation}
respectively. The negative expectation of the last line of Equation~\eqref{eqn:b_derivatives} with respect to $\xvec$ is
\begin{equation}
  \begin{split}
    -E_{\xvec} \left[
      \frac{\partial^2}{\partial \bvec \partial \bvec'} \log p(\bvec \mid \xvec, \mu, \tau^2_b)
    \right]
    & = E_{\xvec} \left[
          \frac{1}{2} \sum_{\freqindex = 1}^\ntimes \frac{I_\freqindex}{e^{\qvec_\freqindex' \bvec}} \qvec_\freqindex' \qvec_\freqindex
          + \Sigma_b^{-1}
        \right] \\
    & = \frac{1}{2} \sum_{\freqindex = 1}^\ntimes E_{\xvec}\left[ \frac{I_\freqindex}{e^{\qvec_\freqindex' \bvec}} \right] \qvec_\freqindex' \qvec_\freqindex
        + \Sigma_b^{-1} \\
    & = \frac{1}{2} \sum_{\freqindex = 1}^\ntimes \qvec_\freqindex' \qvec_\freqindex
        + \Sigma_b^{-1},
  \end{split}
  \label{eqn:fisher_information}
\end{equation}
where the last step follows from the fact that the Whittle likelihood (Equation~\eqref{eqn:whittle_likelihood}) implies that $I_\freqindex \sim f(\omega_\freqindex) \chi^2_1$ for $k = 1$ and $k = \frac{n}{2} + 1$, and $I_\freqindex \sim f(\omega_\freqindex) \chi^2_2 / 2$ otherwise, so that $E_{\xvec}[I_\freqindex / e^{\qvec_\freqindex' \bvec}] = E_{\xvec}[I_\freqindex / f(\omega_\freqindex)] = 1$.

Equation~\eqref{eqn:fisher_information} shows that the metric tensor, $-E_{\xvec} \left[ \frac{\partial^2}{\partial \bvec \partial \bvec'} \log p(\bvec \mid \xvec, \mu, \tau^2_b) \right]$, is constant with respect to $\bvec$. A constant metric tensor corresponds to a special case of RMHMC, avoiding computational complexity required in general. This special case is equivalent to plain HMC with a mass matrix equal to the metric tensor \citep[see][for an explanation of the term mass matrix, and the terms leap frog step size and number of steps used below]{neal2011}. We add an RMHMC step to each iteration of the sampling scheme of RWS12, picking a segment $s \in \{ 1, \ldots, m \}$ uniformly at random, and updating $\bvec_{s, m}$ using RMHMC. The HMC leap frog step size and number of steps used are chosen uniformly at random from $[0.1, 1]$ and $[1, 10]$, respectively.

\subsection{\texorpdfstring{Drawing $\zvec$}{z}}

The mixture component indicators have the conditional probability mass function
\[
  p(z_\siteindex = \componentindex \mid \betavec_\componentindex, \Theta_\componentindex, \xvec^\all) = \frac{
    \pi_\componentindex(\uvec_\siteindex) g_\componentindex(\xvec_\siteindex \mid \Theta_\componentindex)
  }{
    \sum_{\componentindex' = 1}^\ncomponents \pi_{\componentindex'}(\uvec_\siteindex) g_{\componentindex'}(\xvec_\siteindex \mid \Theta_{\componentindex'})
  },
\]
which can be sampled directly for each $\siteindex = 1, 2, \ldots, \nsites$.

\subsection{\texorpdfstring{Drawing $\betavec^\dagger_\componentindex$}{beta star}}

To sample $\betavec^\dagger_\componentindex$ we draw on the work of \citet{rigondurante2020}, who extend the data augmentation scheme of \citet{polsonetal2013}. Let $Z_\componentindex = \{ \siteindex : \siteindex \in \{ 1, 2, \ldots, \nsites \}, z_\siteindex \geq \componentindex \}$, and $U^\dagger_\componentindex$ be the matrix with rows given by $\uvec^{\dagger\prime}_\siteindex$ for $\siteindex \in Z_\componentindex$. The model is augmented with variables $\eta_{\siteindex, \componentindex}$ so that
\begin{align*}
  (\eta_{\siteindex, \componentindex} \mid \betavec^\dagger_\componentindex)
  & \sim \text{PG}(1, \uvec^{\dagger\prime}_\siteindex \betavec^\dagger_\componentindex) \text{ for } \siteindex \in Z_\componentindex, \\
  (\betavec^\dagger_\componentindex \mid \etavec_\componentindex, \tau_\componentindex, \zvec)
  & \sim \N(\mvec_\componentindex, \Sigma^\text{PG}_\componentindex),
\end{align*}
where $\text{PG}$ denotes the P\'{o}lya-Gamma distribution \citep[see][]{polsonetal2013},
\begin{align*}
  \Sigma^\text{PG}_\componentindex
  & = (U^{\dagger \prime}_\componentindex \diag(\etavec_\componentindex) U^\dagger_\componentindex + \Sigma_{\beta^\dagger}^{-1})^{-1}, \\
  \mvec_\componentindex
  & = \Sigma^\text{PG}_\componentindex (U^{\dagger \prime}\kappavec_\componentindex + \Sigma_{\beta^\dagger}^{-1} \muvec^\dagger),
\end{align*}
$\etavec_\componentindex = (\eta_{\siteindex, \componentindex})_{\siteindex \in Z_\componentindex}'$, $\kappa_{\siteindex, \componentindex} = \ind(z_\siteindex = \componentindex) - 0.5$, $\ind(z_\siteindex = \componentindex)$ is 1 if $z_\siteindex = \componentindex$ and 0 otherwise, $\kappavec_\componentindex = (\kappa_{\siteindex, \componentindex})_{\siteindex \in Z_\componentindex}'$, $\muvec^\dagger = (\muvec_\beta', \bm{0}_B')'$, and $\Sigma_{\beta^\dagger} = \begin{pmatrix} \Sigma_\beta & 0 \\ 0 & \tau^2_\componentindex I_B \end{pmatrix}$. Sampling then proceeds for each $k \in \{ 1, 2, \ldots, \ncomponents \}$ by sampling $\eta_{\siteindex, \componentindex}$ for each $\siteindex \in Z_\componentindex$, then sampling $\betavec^\dagger_\componentindex$.

\subsection{\texorpdfstring{Drawing $\tau_\componentindex$}{tau}}

We follow \citet{wand2012}, expanding the half-$t$ prior by augmenting the model with a latent variable $a_\componentindex$, using the hierarchical structure
\[
  (\tau_\componentindex^2 \mid a_\componentindex) \sim \mathrm{IG}\left( \frac{\nu_\tau}{2}, \frac{\nu_\tau}{a_\componentindex} \right)\text{, }
  a_\componentindex \sim \mathrm{IG}\left( \frac{1}{2}, \frac{1}{A_\tau^2} \right),
\]
so that the full conditional distributions are
\begin{align*}
  (a_\componentindex \mid \tau_\componentindex^2)
  & \sim \mathrm{IG}\left( \frac{\nu_\tau + 1}{2}, \frac{\nu_\tau}{\tau_\componentindex^2} + \frac{1}{A_\tau^2} \right), \text{ and} \\[2em]
  (\tau_\componentindex^2 \mid \betavec^\dagger_\componentindex, a_\componentindex)
  & \sim \mathrm{IG}\left( \frac{\nu_\tau + R}{2}, \frac{\betavec^{\mathrm{GP}\prime}_\componentindex\betavec^\mathrm{GP}_\componentindex}{2} + \frac{\nu_\tau}{a_\componentindex} \right).
\end{align*}
The sampling scheme therefore proceeds by first sampling from $(a_\componentindex^{[l + 1]} \mid \tau_\componentindex^{2[l]})$, then $(\tau_\componentindex^{2[l + 1]} \mid a_\componentindex^{[l + 1]}, \betavec^{\mathrm{GP}[l + 1]})$.

\renewcommand{\theequation}{B.\arabic{equation}}
\setcounter{equation}{0}

\section{The structure of \texorpdfstring{$\Lambda$}{Lambda}}
\label{sec:appendix_lambda_derivation}

This appendix presents the derivation for the structure of the matrix $\Lambda$ of Section~\ref{sec:model_missing_values}. The matrix $\Lambda$ has entries
\begin{align*}
  \Lambda_{\timeindex_1, \timeindex_2}
  & = \sum_{\freqindex = 1}^\ntimes
      \frac{1}{f(\omega_\freqindex)}
      V_{\timeindex_1, \freqindex}
      V_{\freqindex, \timeindex_2}^* \\
  & = \frac{1}{\ntimes} \sum_{\freqindex = 1}^\ntimes
      \frac{1}{f(\omega_\freqindex)}
      e^{-2\pi i (\timeindex_1 - \timeindex_2) \omega_\freqindex} \\
  & \equiv
    \lambda_{\timeindex_1 - \timeindex_2},
\end{align*}
where the notation in the final line expresses the fact that $\Lambda_{\timeindex_1, \timeindex_2}$ depends only on $\timeindex_1 - \timeindex_2$. The matrix $\Lambda$ is therefore Toeplitz. Letting $\lambdavec = (\lambda_0, \lambda_1, \ldots, \lambda_{\ntimes - 1})'$, we have $\lambdavec = \frac{1}{\sqrt{n}} V' \rvec$, that is, $\lambdavec$ is the DFT of $\rvec$ divided by $\sqrt{n}$. In fact, $\lambdavec$ is real valued, as
\begin{align*}
  \lambda_\timeindex
  & = \frac{1}{\ntimes} \sum_{\freqindex = 1}^\ntimes
      \frac{1}{f(\omega_\freqindex)}
      e^{-2\pi i \timeindex \omega_\freqindex} \\
  & = \frac{1}{\ntimes} \sum_{\freqindex = 1}^\ntimes
      \frac{1}{f(\omega_\freqindex)}
      [\cos(2\pi \timeindex \omega_\freqindex) - i\sin(2\pi \timeindex \omega_\freqindex)],
\end{align*}
and the fact that $f(\omega_\freqindex) = f(\omega_{\ntimes - \freqindex + 2})$ and $-\sin(2\pi \timeindex \omega_\freqindex) = \sin(2\pi \timeindex \omega_{\ntimes - \freqindex + 2})$ ensure that the complex parts of the sum cancel. Finally, $\lambda_\timeindex = \lambda_{\ntimes - \timeindex}$, so that $\Lambda$ has a symmetric circulant structure.

\end{document}